\def\be{\begin{equation}}
\def\eq{\end{equation}}
\newcommand{\bea}{\begin{eqnarray}}
\newcommand{\eea}{\end{eqnarray}}
\newcommand{\ba}{\begin{eqnarray*}}
\newcommand{\ea}{\end{eqnarray*}}
\newcommand{\nn}{\nonumber}

\documentclass[aps,prd,preprintnumbers,showpacs]{revtex4-1}

\usepackage{amsmath}
\usepackage{amsfonts} 
\usepackage{amssymb}
\usepackage{xcolor}
\usepackage{graphicx}
\usepackage{feynmf}
\usepackage{dcolumn}
\usepackage{bm}
\usepackage{verbatim}
\usepackage{float}
\usepackage{slashed}
\usepackage{appendix}

\DeclareMathOperator{\erf}{Erf}
\DeclareMathOperator{\har}{H}
\DeclareMathOperator{\ghf}{F_{2}^{2}}

\begin{document}

\preprint{EFI 09-17}

\title{Bound states and fermiophobic Unparticle oblique corrections to the photon}

\author{Arun M. Thalapillil}
\email{madhav@uchicago.edu}
\affiliation{Enrico Fermi Institute and Department of Physics, University of Chicago, 5640 South Ellis Avenue, Chicago, IL 60637\\}

\date{\today}
\date{\today}

\begin{abstract}
We study the effects of fermiophobic scalar/pseudo-scalar oblique corrections on bound state energy levels in muonic atoms. To make the treatment sufficiently general, while including ordinary scalar and axion-like pseudo-scalar fields as special cases, we consider Unparticle scalar/pseudo-scalar operators with couplings predominantly to photons. We derive the relevant vacuum polarization functions and comment on the functional forms of the Unparticle Uehling potentials for various scaling dimensions in the point nucleus and finite nucleus approximations. It is estimated that for an infra-red fixed point near the scale of electroweak symmetry breaking, in the low TeV range, and natural values for the model parameters, the energy shifts in the low-lying muonic lead transitions are typically of the order of a few times 0.1 eV to a few times 0.01 eV. The energy level structure of the Unparticle Uehling shifts are inferred using general methods for the scalar and pseudo-scalar cases and it is shown that the two cases contribute to the energy shifts with the same sign.  It is shown that this conclusion is not changed even when scale invariance is broken and is in fact relatively insensitive to the scale at which it is broken. It is pointed out nevertheless that the estimated magnitude of the Unparticle Uehling shift (based on some \textit{natural} values for the model parameters) is a factor of $1000-10000$ below the discrepancy in QED/nuclear theory and precision muonic lead spectroscopy from about two decades ago. We briefly comment on scenarios where the Unparticle induced energy shift, if it exists, may be experimentally measurable. One possibility in this direction is if the UV-sector, from which the Unparticle sector arises, has a large number of fermions. Comments are also made on the possibility of further studying muonic atoms, as a probe for beyond-standard-model physics, in the context of forthcoming experiments, such as those probing lepton flavor violation through coherent muon-electron conversions. For completeness we explore some of the astrophysical and cosmological consequences of a fermiophobic scalar/pseudo-scalar Unparticle sector. In the fermiophobic context we also estimate a minimum value for the conformal invariance breaking scale. 
\end{abstract}
\pacs{12.60.-i,12.90.+b,14.80.-j}

\maketitle

\begin{section}{Introduction}
\par
Recently there has been much interest in the possibility of a scale invariant hidden sector that couples to the standard model (SM)~\cite{{Georgi:2007si}, {Banks:1981nn}}. The operators in such a theory have been referred to as ``Unparticles" to emphasize their generally \textit{non-integral} scaling dimensions. In addition to an n-body phase space resembling that of a non-integral number of massless particles, the correlation functions of the Unparticle operators have interesting \textit{non-trivial phases}~\cite{Georgi:2007si}. The above properties make this sector distinct from other beyond-SM extensions. 
\par
A model with the above properties was proposed by Banks and Zaks (BZ)~\cite{Banks:1981nn}. Their theory was a vector-like $SU(3)$ gauge theory with $N_{\text{f}}$ fermions in the fundamental representation. It was found that for a particular range of $N_{\text{f}}$ the $\beta$ function at lowest order is negative ($\beta_{1}(N_{\text{f}})<0$) while the contribution to the $\beta$ function at the next order is positive ($\beta_{2}(N_{\text{f}})>0$)  in
\ba
\beta(g)\simeq~\beta_{1}(N_{\text{f}})\,\frac{g^3}{16\pi^2}+\beta_{2}(N_{\text{f}})\,\frac{g^5}{(16\pi^2)^2}+\,\ldots
\ea
This would mean that as the theory flows to lower energies the small coupling constant $g$ would grow until it hits an infra-red fixed point where
\ba
\beta(g^{*})\simeq~0
\ea
The theory is scale-invariant below this scale and the description in terms of  the Banks-Zaks fields at high energies is replaced by one in terms of composite particles of a strongly-coupled scale-invariant theory. These composite particles may be identified with the Unparticles~\cite{Georgi:2007si}. We are interested in exploring the case of a scale invariant \textit{fermiophobic Unparticle sector} that couples only with the massless SM gauge bosons, specifically with a substantial coupling only to the photon. As we shall see a fermiophobic sector might be able to avoid certain constraints compared to a fermiophilic sector. Also, independent of considerations in our study there is great interest in a fermiophobic scalar sector in the context of electroweak symmetry breaking and the Higgs mechanism~\cite{landsberg_cite}. Our main focus will be on a fermiophobic sector that couples predominantly to photons and the effects of the induced oblique corrections on atomic energy levels in muonic atoms.
\par
If one assumes that the scale invariant Unparticle sector is also conformally invariant, then the scaling dimensions ($j$) of the gauge invariant primary Unparticle operators are tightly constrained by requirements of conformal invariance. For an operator in the $(l_{1},\,l_{2})$ Lorentz spin representation the constraints are~\cite{{Mack:1975je},{Nakayama:2007qu},{Grinstein:2008qk}}
\ba
j\,\geq\,l_{1}+l_{2}+2-\delta_{l_{1}\times l_{2},0}
\ea
These bounds translate to
\ba
j_{\mathcal{U}}\geq~1~;~~j_{\mathcal{U}_{\text{\tiny{f}}}}\geq~3/2~;~~j_{\mathcal{U}_{\text{\tiny{V}}}}~\geq~3
\ea
for the scalar, fermion and vector Unparticle operators respectively. These conditions are referred to as Mack's unitarity criteria~\cite{Mack:1975je}. We will impose these constraints on the Unparticle operators we work with in the present study. Although it is technically possible for a quantum field theory to be scale invariant but not conformally-invariant, examples are rare~\cite{Gross:1970tb}.
\par
The Unparticle operators in the limit of exact scale invariance may be considered as a sum over resonances having a continuous mass distribution with no mass gap~\cite{Stephanov:2007ry}. The K\"{a}ll\'en-Lehmann spectral density of  the Unparticle operator takes the form~\cite{Georgi:2007si}
\ba
\rho(s^2)~\propto~(s^{2})^{j-2}
\ea
This implies that for $j\geq 2$ the theory becomes very UV sensitive and may cause singular behavior~\cite{Rajaraman:2008qt}. This suggests that we only consider operators with $j<2$. This is in conflict with the requirements of Mack's unitarity for primary, vector Unparticle operators which require $j\geq 3$~\cite{Rajaraman:2008qt}. Usually in any Unparticle model with couplings to SM fields there must also be additional \textit{contact terms} between SM operators as first pointed out in~\cite{Grinstein:2008qk}. They are generated when we integrate out the ultra-heavy UV fields (BZ fields for example). Without some fine-tuning these terms generally dominate over the SM-Unparticle couplings and render Unparticle phenomenology to be of lesser interest. In the case of scalar/pseudo-scalar Unparticles where the Mack's unitarity constraint is $j\geq 1$ and the scaling dimension may be very close to unity these contact interactions may nevertheless be of lesser importance~\cite{Grinstein:2008qk}.
\par
Apart from simplicity the arguments above will be our motivation for considering scalar/pseudo-scalar Unparticles ($\mathcal{U}$) in our study. In the subsequent analysis we will use $\mathcal{U}$ to label a generic scalar/pseudo-scalar Unparticle, $\mathcal{O}$ for an Unparticle scalar and $\tilde{\mathcal{O}}$ for an Unparticle pseudo-scalar. Also, we label an ordinary scalar by $\phi$, pseudo-scalar by $\tilde{\phi}$ and a generic scalar/pseudo-scalar operator by $\Phi$. The energy scale of the infra-red fixed point (where the theory becomes scale invariant) is usually denoted by $\Lambda_{\mathcal{U}}$ and that at which scale invariance broken by the parameter $\mu$. We will assume that the Unparticle scale is in the low TeV range near the scale of electroweak-symmetry breaking, i.e., $\Lambda_{\mathcal{U}}\sim\, v$. Specifically we will interpret this to mean $\Lambda_{\mathcal{U}}\in[\sim246\,\text{GeV},\,\mathcal{O}(1)\,\text{TeV}]$. There are stringent constraints on
\ba
p\bar{p}\rightarrow h_{f}\rightarrow\,\gamma\gamma+\,X
\ea
where $h_{f}$ is a fermiophobic Higgs, from the CDF and D0 collaborations at the Tevatron~\cite{:2008it}. In our case the coupling of the fermiophobic scalar/pseudo-scalar sector to the fermions and the heavy gauge bosons ($Z^{0},\,W^{\pm}$) are assumed to be very small or close to zero. Also, the coupling to the $SU(3)_{c}$ colored massless gluons is assumed to be much smaller than the coupling to the photon. In this case the interaction scale may be in the $\Lambda_{\mathcal{U}}\sim\, v$ range we consider above, in the low TeV regime, and still be consistent with the collider constraints~\cite{Feng:2008ae}. We will estimate a lower bound for the scale of $\mu$ in the fermiophobic Unparticle case later in section III.
\par
Based on all the above arguments we will therefore consider Unparticle scalars and pseudo-scalars coupling to photons with an interaction scale $\Lambda_{\mathcal{U}}\sim\, v$ and satisfying
\be
1\,\leq~j_{\mathcal{U}}\,<~2
\eq
\par
Now if one considers two representative fermiophobic and fermiophilic effective interaction terms of the form 
\ba
\Lambda^{-j}_{\gamma}\,\mathcal{U}F_{\alpha\beta}F^{\alpha\beta}~~;~~~~\Lambda^{1-j}_{\psi}\,\mathcal{U}_{\text{\tiny{V}}}^{\alpha}\bar{\Psi}\gamma_{\alpha}\Psi
\ea
it may be argued based on scaling arguments and dimensional analysis~\cite{Bander:2007nd} that in general
\ba
\Lambda_{\psi}~>~\Lambda_{\gamma}
\ea
This seems to imply, for the relevant values of $j$, that the coupling to the gauge bosons may lead to larger effects than the corresponding fermiophilic case~\cite{Bander:2007nd}. Although the true niche for probing such effects may be in high energy colliders (see for example~\cite{{Rajaraman:2008qt}, {Bander:2007nd},{Cheung:2007zza},{Luo:2007bq},{Fox:2007sy},{Feng:2008ae},{Barger:2008jt},{Cheung:2008xu}} and references therein), especially when the interaction scale is much higher than $\mathcal{O}(1)\,\text{TeV}$, it is still interesting to explore the effects of a fermiophobic coupling in low-energy experiments. In the fermiophilic case for instance one can put interesting bounds on the Unparticle sector from atomic parity violation~\cite{Bhattacharyya:2007pi}. 
\par
We are specifically interested in fermiophobic Unparticle contributions to muonic atom transitions. Although in this case, unlike atomic parity violation in the fermiophilic case~\cite{Bhattacharyya:2007pi}, no symmetry is violated there could be atomic systems or regions in parameter space where the effect may be measurable. Also, if the fermiophobic Unparticle sector is such that it has substantial couplings only to photons and no other gauge bosons, the muonic atom transitions can play a unique role in constraining it. 
\par
We chose to study fermiophobic Unparticle contributions to muonic lead transitions. It will be seen in this specific case that for \textit{natural values of the model parameters} the estimated Unparticle induced energy shifts in low-lying muonic-lead atomic transitions may be of the \textit{order of a few 0.1-0.01 \text{eV}} which is comparable to the bound state QED corrections from light-by-light scattering and the fourth order Lamb shift to higher orbital angular momentum transitions respectively. We will comment on cases where this value may be enhanced or measurable, for instance when there is a very large fermion \textit{multiplicity} in the UV sector.
\par
Another point is that since oblique corrections may potentially involve new ``heavy" particles running in the loops we may expect them to be a probe for Unparticles within a wide range of $\mu$ values. We will discuss this aspect, in the context of the Unparticle Uehling energy shifts, in more detail in section III. There is also the motivation that many of the constraints on the couplings of an Unparticle to SM fields~\cite{Freitas:2007ip} are more relaxed in the case of a fermiophobic sector. We will discuss the case of broken scale invariance and the requirements for evading astrophysical and cosmological constraints later in our study. 
\par
The scalar/pseudo-scalar Unparticle propagator in the limit of exact scale invariance is \cite{Georgi:2007si}
\be
\int d^{4}x~e^{iqx}\langle0\vert T \left(\mathcal{O}_{\mathcal{U}}(x)\mathcal{O}_{\mathcal{U}}(0)\right)\vert0\rangle=\frac{i\,A_{j}}{2\,\sin(j\pi)}\left[-\left(q^2+i\epsilon\right)\right]^{j-2}
\label{muzprop}
\eq
where
\ba
A_{j}=\frac{16\pi^{5/2}\Gamma(j+1/2)}{(2\pi)^{2j}\Gamma(j-1)\Gamma(2j)}
\label{aj}
\ea
with $j$ being the scaling dimension of the Unparticle operator. Note that since $j$ is in general a non-integer there is a non-trivial phase $e^{i(2-j)\pi}$ associated with the Unparticle propagator. Also note that in the limit of $j\rightarrow\,1^{+}$ there is \textit{no singular behavior} and we recover the ordinary scalar/pseudo-scalar propagator. The propagator coefficient $A_{j}/(2\sin j\pi)$ has unit magnitude at $j=1$ and diverges very close to $j=2$ where the theory is very UV sensitive. In the rest of the region its magnitude is below $1$.
\par
The case of broken scale invariance may be parametrized by introducing an effective mass gap $\mu$ in the spectral decomposition leading to the scalar/pseudo-scalar Unparticle propagator~\cite{{Fox:2007sy},{Barger:2008jt}}
\be
\int d^{4}x~e^{iqx}\langle0\vert T \left(\mathcal{O}_{\mathcal{U}}(x)\mathcal{O}_{\mathcal{U}}(0)\right)\vert0\rangle=\frac{i\,A_{j}}{2\,\sin(j\pi)}\left[-\left(q^2-\mu^{2}+i\epsilon\right)\right]^{j-2}
\label{muprop}
\eq
$\mu$ may be thought of as the scale at which conformal invariance is effectively broken.
\par
Let the fermiophobic Unparticle fields be gauge singlets under the SM gauge group. Then the coupling of the scalar Unparticle to two photons may be incorporated~\cite{Georgi:2007si} by a term in the effective low-energy action
\be
\mathcal{S}^{\text{\tiny{\,eff.}}}_{\text{\tiny{S}}}=\int d^{4}x~\frac{c}{4\,\Lambda^{j}_{\gamma}}\,\mathcal{O}\,F_{\mu\nu}\,F^{\mu\nu}\,+\,\ldots
\label{slag}
\eq
Here $\ldots$ denotes terms that are suppressed by higher powers of the relevant energy scale and which have been ignored. $\Lambda_{\gamma}$ is a scale relevant to the $\gamma-\mathcal{O}-\gamma$ coupling derived from the fundamental Unparticle scale $\Lambda_{\mathcal{U}}$. The coefficient $c$ is assumed to be an $\mathcal{O}(1)$ constant factor. Note that when the Unparticle sector is being generated from a UV theory (such as the BZ theory) the coefficient of the above operator goes like~\cite{Georgi:2007si}
\ba
\sim~\mathcal{O}(1)\,\left(\frac{\Lambda_{\mathcal{U}}}{M_{\text{\tiny{UV}}}}\right)^{d_{\text{\tiny{UV}}}}
\ea
and is in general not of $\mathcal{O}(1)$. Here $M_{\text{\tiny{UV}}}$ is the scale of the UV physics and $d_{\text{\tiny{UV}}}$ is the dimension of the UV operator coupling to $F_{\mu\nu}\,F^{\mu\nu}$. But if one assumes, for example, that the scalar Unparticle operators $\mathcal{O}_{i}$ are being generated by the confinement ($\,\langle\bar{\Psi}_{i}\Psi_{i}\rangle\rightarrow \mathcal{O}_{i}$) of fermions ($\Psi_{i}$) from the UV sector (in this case $d_{\text{\tiny{UV}}}=3$), conservatively, the ``effective" coefficient 
\ba
c~\simeq~\sum_{i=1}^{N_{f}}\,c_{i}~\sim~N_{f}~\mathcal{O}(1)\,\left(\frac{\Lambda_{\mathcal{U}}}{M_{\text{\tiny{UV}}}}\right)^{d_{\text{\tiny{UV}}}}
\ea
could \textit{naturally} be of $\mathcal{O}(1)$. The reason is that theoretically the number of fermions $N_{f}$ in the UV theory is permitted to be large and is constrained only by the requirement of conformal invariance at $g^{*}$. For example
\ba
\frac{33}{2}~\gtrsim~N_{f}~\gtrsim~\frac{306}{38}
\ea
in the BZ theory~\cite{Banks:1981nn} and in a recent Technicolor inspired $SU(N_{T})\times SU(N_{U})$ model by F. Sannino and R. Zwicky~\cite{Sannino:2008nv}
\ba
\frac{11}{\gamma^{*}+2}\,N_{T}~\gtrsim~N_{f}~\gtrsim~\frac{11}{\gamma^{*}+2}\,N_{U}+2
\ea
where the critical anomalous dimension satisfies the unitarity bound $\gamma^{*}\leq 2$. Thus we will interpret the effective interaction in Eq.\,(\ref{slag}) as modeling the effects of various possible Unparticle scalar operators ($\mathcal{O}\approx\,\sum_{i=1}^{N_{f}}\,\mathcal{O}_{i}$), in a semi-realistic model, and take the coefficent $c\sim\mathcal{O}(1)$ without loss of generality. But it is nevertheless important to keep in mind that apart from notions of \textit{naturalness} nothing excludes a larger value for the coupling, for instance if there is very large fermion \textit{multiplicity} (ie. very large $N_{\text{\tiny{F}}}$) in the hidden sector. QCD-like models with a possibly large number of colors ($N_{\text{\tiny{C}}}$) and fermion flavors ($N_{\text{\tiny{F}}}$) is not uncommon, for example, in many string-inspired models~\cite{Sakai:2004cn}.

\par
Similarly the coupling of a pseudo-scalar Unparticle to two photons may be modeled by 
\be
\mathcal{S}^{\text{\tiny{\,eff.}}}_{\text{\tiny{PS}}}=\int d^{4}x~\frac{b}{4\,\tilde{\Lambda}^{j}_{\gamma}}\,\mathcal{\widetilde{O}}\,F^{\mu\nu}\,\widetilde{F}_{\mu\nu}\,+\,\ldots
\label{pslag}
\eq
where
\ba
\widetilde{F}_{\mu\nu}=\frac{1}{2}\,\epsilon_{\alpha\beta\mu\nu}\,F^{\alpha\beta}
\ea
$\tilde{\Lambda}_{\gamma}$ is a scale relevant to the $\gamma-\tilde{\mathcal{O}}-\gamma$ coupling and the constant $b$ is assumed to be of $\mathcal{O}(1)$. All the assumptions in the scalar case again hold here. The above may be compared to the chiral anomaly induced $\gamma-\pi^{0}-\gamma$ Wess-Zumino-Witten coupling  
\be
\mathcal{S}_{\pi \gamma \gamma}=\int d^{4}x~\frac{-N_{c}\,e^{2}}{48\pi^{2}\,f_{\pi}}\,\pi^{0}\,F_{\mu\nu}\,\tilde{F}^{\mu\nu}
\label{wzw}
\eq
leading to
\ba
\Gamma_{\mu}\xrightarrow{q_{\pi}\rightarrow0}\,-\frac{ie^{2}}{4\pi^2 f_{\pi}}\,\epsilon_{\alpha\beta\rho\sigma}k^{\rho}l^{\sigma}
\ea
valid in the soft pion limit (Here $N_{c}$ is the number of `colors', $k^{\rho}$ and $l^{\sigma}$ are the four-momenta of the two photons) and the $\gamma^{*}-\gamma-\tilde{\phi}$ vertex (here $\tilde{\phi} $ is a pseudo-scalar meson like $\pi^{0},\,\eta~\text{or}~\eta^{'}$)
\ba
\Gamma_{\mu}\xrightarrow{Q^{2}\rightarrow\pm\infty}\,-ie^{2}\,\left(\frac{2f_{\pi}}{Q^{2}}\right)\,\epsilon_{\mu\nu\rho\sigma}p^{\nu}\epsilon^{\rho}q^{\sigma}
\ea
proposed in~\cite{Brodsky:1981rp} for $Q^{2}=-q^{2}\rightarrow\pm\infty$ and recently considered in~\cite{Rosner:2009bp} in the context of meson-photon transition form factors in the charmonium energy range. In the above expression $q^{\sigma}$ is the four-momentum of the off-shell photon $\gamma^{*}$, $p^{\nu}$ is the four-momentum of the pseudo-scalar meson $\tilde{\phi}$, $\epsilon^{\rho}$ is the polarization vector of the outgoing on-shell photon $\gamma$ and $f_{\pi}\simeq93\,\text{MeV}$ is the pion constant.  
\par
The corrections to muonic atom transitions due to photon vacuum polarization are induced by diagrams like those in Fig.\,\ref{unvp}. It is our aim to estimate the magnitude of such contributions to low-lying muonic atom levels. We mention, as an aside, that processes such as those in Fig.\,\ref{unvp} for the case of low-energy QCD (where $\mathcal{U}(k)$ is now again a pseudo-scalar meson like $\pi^{0}\,\text{or}~\eta$) contribute for instance to muon (g-2). Using Chiral perturbation theory based on Eq.\,(\ref{wzw}) and $\omega$ vector-meson dominance, in the relevant range $\sqrt{s}<\,0.6\,\,\text{GeV}$, a diagram such as that in Fig.\,\ref{unvp} for the $\pi^{0}$ is expected to contribute~\cite{Hagiwara:2003da}
\ba
a_{\mu}(\pi^{0}\gamma,\sqrt{s}<\,0.6\,\,\text{GeV})=(0.13\pm0.01)\times\,10^{-10}
\ea
A similar contribution from the fermiophobic Unparticle sector, for typical model parameter values, is expected to be very small and within experimental limits, since the QED Kernel $K(s)$~\cite{Brodsky:1967sr} has a steep cut-off around $\mathcal{O}(1)\,\text{GeV}$.
\par
Let us now proceed to calculate the vacuum polarization functions and estimate the induced effective Uehling potentials for various cases.
\end{section}


\begin{section}{Oblique corrections and the Unparticle Uehling potential}
\par
We are primarily interested in possible oblique corrections to the photon propagator due to scalar/pseudo-scalar Unparticles as shown in Fig.\,\ref{unvp}. If they exist we expect such vacuum polarizations to modify the photon propagator over the usual SM corrections. These scalar/pseudo-scalar oblique corrections can show up potentially as very tiny energy shifts in muonic atom energy levels or in the anomalous magnetic moment of the muon. As we have mentioned previously our main focus will be on possible corrections to the atomic transitions in muonic atoms. We will first consider the case of perfect scale invariance $\mu\rightarrow 0$. The case when $\mu\neq0$ will be discussed in section III after we have determined a minimum value for $\mu$ in the fermiophobic case as dictated by astrophysical and cosmological constraints.

\begin{figure}
\includegraphics[width=9.5cm,angle=0]{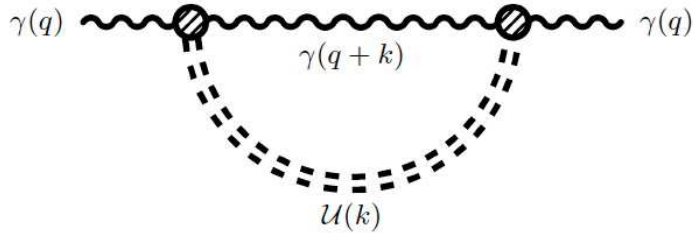}
\caption{Photon vacuum polarization by an Unparticle scalar or pseudo-scalar with an arbitrary scaling dimension $j$. The vertices are shown with blobs to indicate that they are effective couplings coming from an effective low-energy action of the form (\ref{slag}) or (\ref{pslag}).}
\label{unvp}
\end{figure}
\par
Using the Feynman rule for Eq.\,(\ref{slag}) (see for example~\cite{Cheung:2007ap}) the scalar Unparticle contribution to the photon polarization tensor is
\ba
i\,\Pi_{\mathcal{O}}^{\mu\nu}=\,-\frac{c^2A_{j}}{2\Lambda_{\gamma}^{2j}\sin(j\pi)}\int \frac{d^{4}k}{(2\pi)^{4}}\frac{\left[q\cdot(q+k)g^{\alpha\mu}-q^{\alpha}(q+k)^{\mu}\right]\,\left[q\cdot(q+k)g^{\nu}_{\alpha}-q_{\alpha}(q+k)^{\nu}\right]}{\left[(q+k)^2+i\epsilon\right]\left[-(k^2-\mu^2)+i\epsilon\right]^{2-j}}
\ea
This simplifies to the the expression
\be
i\,\Pi_{\mathcal{O}}^{\mu\nu}=\,i\,\Pi_{\mathcal{O}}(q^2,\mu^2,j)~\left(q^{2}g^{\mu\nu}-q^{\mu}q^{\nu}\right)
\label{uspol}
\eq
with
\be
i\,\Pi_{\mathcal{O}}(q^2,\mu^2,j)=-\frac{c^2A_{j}e^{i(2-j)\pi}\Gamma(3-j)}{2\Lambda_{\gamma}^{2j}\sin(j\pi)\Gamma(2-j)}\int dx\,dy\, \delta(x+y-1) ~y^{1-j}\int\frac{d^{4}l}{(2\pi)^{4}}\frac{q^{2}(x-1)^2+(l^{2}/2)}{\left[l^{2}-\Delta(q^2,\mu^2)+i\epsilon\right]^{3-j}}
\label{spolf}
\eq
where
\ba
\Delta(q^2,\mu^2)=\,x\,(x-1)\,q^{2}+\,y\,\mu^{2}
\ea
We note that Eq.\,(\ref{uspol}) has the expected gauge invariant structure and satisfies the Ward identities. 
\par
For the pseudo-scalar Unparticle case the polarization tensor is
\ba
i\,\Pi_{\widetilde{\mathcal{O}}}^{\mu\nu}=\,-\frac{b^2A_{j}}{2\tilde{\Lambda}_{\gamma}^{2j}\sin(j\pi)}\int \frac{d^{4}k}{(2\pi)^{4}}\frac{\left[\epsilon^{\xi\mu\rho\alpha}\epsilon_{\lambda\alpha\kappa\nu}q_{\rho}(q+k)_{\xi}q^{\lambda}(q+k)^{\kappa}\right]}{\left[(q+k)^2+i\epsilon\right]\left[-(k^2-\mu^2)+i\epsilon\right]^{2-j}}
\ea
This may be simplified to give
\be
i\,\Pi_{\widetilde{\mathcal{O}}}^{\mu\nu}=\,i\,\Pi_{\widetilde{\mathcal{O}}}(q^2,\mu^2,j)~\left(q^{2}g^{\mu\nu}-q^{\mu}q^{\nu}\right)
\label{upspol}
\eq
where
\be
i\,\Pi_{\widetilde{\mathcal{O}}}(q^2,\mu^2,j)=+\frac{b^2A_{j}e^{i(2-j)\pi}\Gamma(3-j)}{2\tilde{\Lambda}_{\gamma}^{2j}\sin(j\pi)\Gamma(2-j)}\int dx\,dy\, \delta(x+y-1) ~y^{1-j}\int\frac{d^{4}l}{(2\pi)^{4}}\frac{(l^{2}/2)}{\left[l^{2}-\tilde{\Delta}(q^2,\mu^2)+i\epsilon\right]^{3-j}}
\label{pspolf}
\eq
and again
\ba
\tilde{\Delta}(q^2,\mu^2)=\,x\,(x-1)\,q^{2}+\,y\,\tilde{\mu}^{2}
\ea
\par
Regularizing the momentum integral in Eq.\,(\ref{spolf}) using dimensional continuation yields
\be
\Pi_{\mathcal{O}}(q^2,\mu^2,j)=-\frac{c^2A_{j}}{32\pi^{2}\Lambda_{\gamma}^{2j}\sin(j\pi)}\int dx\,dy\, \delta(x+y-1) ~y^{1-j}\left[\frac{\Delta(q^2,\mu^2)}{j\left(j-1\right)}\Delta^{j-1}(q^2,\mu^2)+q^{2}(x-1)^{2}\frac{\Delta^{j-1}(q^2,\mu^2)}{\left(j-1\right)}\right]
\label{scalarpi}
\eq
For the pseudo-scalar case, again regularizing via dimensional continuation
\be
\Pi_{\widetilde{\mathcal{O}}}(q^2,\mu^2,j)=+\frac{b^2A_{j}}{32\pi^{2}\tilde{\Lambda}_{\gamma}^{2j}\sin(j\pi)}\int dx\,dy\, \delta(x+y-1) ~y^{1-j}\left[\frac{\tilde{\Delta}(q^2,\mu^2)}{j\left(j-1\right)}\tilde{\Delta}^{j-1}(q^2,\mu^2)\right]
\label{pscalarpi}
\eq
We will address the $j\rightarrow 1 ^{+}$ limit of  the expressions in Eqs.\,(\ref{scalarpi}) and (\ref{pscalarpi}) later.
\par
The Unparticle polarization functions may be renormalized as
\bea
\label{renpten}
\hat{\Pi}_{\mathcal{O}}(q^2,\mu^2,j)&=&\Pi_{\mathcal{O}}(q^2,\mu^2,j)-\Pi_{\mathcal{O}}(0,\mu^2,j)\\ \nn
\hat{\Pi}_{\widetilde{\mathcal{O}}}(q^2,\mu^2,j)&=&\Pi_{\widetilde{\mathcal{O}}}(q^2,\mu^2,j)-\Pi_{\widetilde{\mathcal{O}}}(0,\mu^2,j)
\eea
so that as $q\rightarrow0$ the residue of the photon propagator tends to unity.
\par
Note that the non-trivial Unparticle propagator phase $e^{i(2-j)\pi}$ does not make an appearance in Eqs.\,(\ref{scalarpi}) and (\ref{pscalarpi}). It is found from explicit calculation that the phase is removed during the evaluation of the Euclidean loop integrals. Therefore the Unparticle vacuum polarization tensor does not have a complex phase from the propagator contribution and any imaginary part that $\hat{\Pi}_{\mathcal{U}}(q^2,\mu^2,j)$ picks up subsequently, if at all,  should come from the kinematic region that $q^{2}$ occupies. 
\par
We may likewise calculate the polarization functions for \textit{ordinary} scalars ($\phi$) and pseudo-scalars ($\tilde{\phi}$) with a two-photon coupling. The ordinary pseudo-scalar in this case may be compared to an \textit{axion-like particle} ($a$) coupling to photons
\ba
\mathcal{S}_{a \gamma \gamma}=\int d^{4}x~\frac{g_{a\gamma}}{4\,f_{a}}\,a\,F_{\mu\nu}\,\tilde{F}^{\mu\nu}
\ea
where the scale $f_{a}$ and the pseudo-scalar mass $m_{a}$ are independent of each other. Substituting $j=1$ in Eqs.\,(\ref{spolf}) and (\ref{pspolf}) and integrating over the loop momenta using dimensional regularization we get after an $\overline{MS}$ subtraction
\bea
\label{ordpf}
\Pi_{\phi}(q^2,m_{\phi}^2)&=&~\frac{a^2}{16\pi^{2}\Lambda_{\phi}^{2}}\int dx\,dy\, \delta(x+y-1)\left[\Delta^{'}(q^2,m_{\phi}^2)~\log\frac{\Delta^{'}(q^2,m_{\phi}^2)}{M^{2}}+\,q^{2}(x-1)^2~\log\frac{\Delta^{'}(q^2,m_{\phi}^2)}{M^{2}}\right]\\ \nn
\Pi_{\tilde{\phi}}(q^2,m_{\tilde{\phi}}^2)&=&~-\frac{\tilde{a}^2}{16\pi^{2}\tilde{\Lambda}_{\phi}^{2}}\int dx\,dy\, \delta(x+y-1)\left[\Delta^{''}(q^2,m_{\tilde{\phi}}^2)~\log\frac{\Delta^{''}(q^2,m_{\tilde{\phi}}^2)}{M^{2}}\right]
\eea
Here 
\ba
\Delta^{'}(q^2,m_{\phi}^2)&=&\,x(x-1)\,q^{2}+\,y \,m_{\phi}^{2}\\
\Delta^{''}(q^2,m_{\tilde{\phi}}^2)&=&\,x(x-1)\,q^{2}+\,y \,m_{\tilde{\phi}}^{2}
\ea
and we have used the fact, from Eq.\,(\ref{muzprop}), that
\ba
\frac{A_{j}}{2\sin(j\pi)}&\xrightarrow{j\rightarrow 1^{+}}&~-1\\ 
e^{i(2-j)\pi}&\xrightarrow{j\rightarrow 1^{+}}&~-1
\ea
$M$ is an arbitrary subtraction scale. It must be mentioned that this dependence on $M$ will be removed when we explicitly introduce other higher order interactions in the effective Lagrangian. A natural choice is to take $M\simeq\,(\Lambda_{\phi},\,\Lambda_{\tilde{\phi}})\sim\,v$, the energy scale of the relevant interaction. The choice would also ensure that in the $m_{\Phi}\rightarrow 0$ limit we can retain to good approximation, as far as numerical computations are concerned, just the lowest order terms in the effective Lagrangian. This is because with this choice, for the range of $q$ we are interested in (specifically $q\lesssim m_{\mu^{-}}$), we have $q^{2}\log(q^{2}/M^{2})\gg\,q^{2}$. For instance in the analogous case in QCD chiral perturbation theory (ChPT) a choice of $M\simeq\Lambda_{\text{\tiny{QCD}}}\sim\,1\,\text{GeV}$ is appropriate and one can keep to lowest order only the so called chiral logarithm term to estimate some of the low-energy QCD effects (see for example~\cite{Manohar:1995xr} and references cited therein). With this tacitly assumed we proceed to analyze the functional forms of the Uehling potentials in the cases of interest.
\par
If one is interested in bound states the modified electromagnetic four-potential $A^{'}_{\mu}$, due to the vacuum polarizations, is given by 
\bea
\label{potFT}
A^{'}_{\mu}(x)&=&\int~\frac{d^{4}q}{(2\pi)^{4}}~e^{-iq\cdot x}\left[1-\hat{\Pi}(q^2)\right]^{-1}~\mathcal{G}^{\text{\tiny{photon}}}_{\mu\nu}(q^{2})\,\mathcal{J}^{\nu}_{\text{\tiny{source}}}(q)\\\nn
&\simeq&\int~\frac{d^{4}q}{(2\pi)^{4}}~e^{-iq\cdot x}\left[1+\hat{\Pi}(q^2)\right]~A^{(0)}_{\mu}(q)
\eea
where $i\,\mathcal{G}^{\text{\tiny{photon}}}_{\mu\nu}(q^{2})$ is the photon propagator, $\mathcal{J}^{\nu}_{\text{\tiny{source}}}(q)$ is the source four-current in momentum space and $A^{(0)}_{\mu}(q)$ is the four-vector potential without any vacuum polarization corrections. As is well known $1-\hat{\Pi}(q^2)$ acts like a dielectric constant for vacuum
\ba
1-\hat{\Pi}(q^2)\sim\frac{\epsilon(\omega)}{\epsilon_{0}}
\ea
Thus, the imaginary part of the polarization tensor corresponds to the vacuum becoming absorptive. 
In general 
\ba
\hat{\Pi}(q^2)=\hat{\Pi}_{\text{\tiny{SM}}}(q^2)+\hat{\Pi}_{\mathcal{U}}(q^2,\mu^2,j)+\hat{\Pi}_{\text{\tiny{Other}}}(q^{2})
\ea
In our analysis we are interested in the corrections solely due to the Unparticle contribution $\Pi_{\mathcal{U}}(q^2,\mu^2,j)$. We noted previously that the non-trivial Unparticle propagator phase $e^{i(2-j)\pi}$ is effectively cancelled during the evaluation of the loop integrals. From Eqs.\,(\ref{scalarpi}), (\ref{pscalarpi}) and (\ref{ordpf}) we observe that for
\ba
x\,(x-1)\,q^{2}+\,y\,\mu^{2}~<~0~&\longmapsto&~\Im m\left[\Pi_{\mathcal{U}}(q^2,\mu^2,j)\right]\neq~0\\
x(x-1)\,q^{2}+\,y \,m_{\Phi}^{2}~<~0~&\longmapsto&~\Im m\left[\Pi_{\Phi}(q^2,m_{\Phi}^{2},j)\right]\neq~0
\ea
and there is a \textit{branch-cut} starting at $q^2=\mu^2$ and $q^2=m_{\Phi}^{2}$ in analogy with $e^{+}-e^{-}$ vacuum polarization in QED. We will approximate an atomic nucleus of charge $Z$ initially as a static, point-source. For a static source of the electromagnetic field the momentum transfer is space-like ($q^2< 0$) and the polarization tensor $\hat{\Pi}_{\mathcal{U}}(-\vec{q}^{\,2},\mu^2,j)$ is real which would imply that there is no absorption in vacuum under these circumstances. 
\par
Approximating an atomic nucleus of charge $Z$ as a static, point-source the electromagnetic 4-current
may be calculated in momentum space to be
\ba
\mathcal{J}^{\nu}_{\text{\tiny{source}}}(q)\simeq-Ze\,\delta^{\nu 0}\,\delta(q^{0})
\ea
In our convention the charge $e$ is intrinsically negative. Then the expression for the modified potential becomes
\be
A^{'\,\text{\tiny{point}}}_{0}(\vec{r})\simeq~-Ze\int~\frac{d^{3}\vec{q}}{(2\pi)^{3}}~e^{i\vec{q}\cdot \vec{r}}\left[1+\hat{\Pi}(-\vec{q}^{\,2}, \mu^2, j)\right]~
\mathcal{G}^{\text{\tiny{photon}}}_{00}(-\vec{q}^{\,2})
\eq
\par
The $\Pi_{\mathcal{U}}(q^2,\mu^2,j)$ contributes to a Uehling potential (in a way similar to $e^{+}-e^{-}$ vacuum polarization in QED) and gives
\be
V^{\text{\tiny{point}}}_{\mathcal{U}}(r, j)\simeq~-Ze\int~\frac{d^{3}\vec{q}}{(2\pi)^{3}}~e^{i\vec{q}\cdot \vec{r}}~\hat{\Pi}_{\mathcal{U}}(-\vec{q}^{\,2}, \mu^2, j)~
\mathcal{G}^{\text{\tiny{photon}}}_{00}(-\vec{q}^{\,2})
\label{pot3FT}
\eq
This is the correction to the electromagnetic potential due to fermiophobic, scalar/pseudo-scalar Unparticle vacuum polarizations.

\begin{figure}
\includegraphics[width=6.5cm,angle=0]{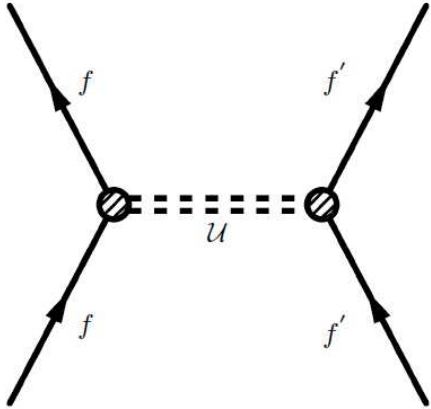}
\caption{Direct scalar/pseudo-scalar Unparticle exchange in a fermiophilic scenario where the Unparticles have substantial fermion couplings. These possibilities are heavily constrained, especially when $\mu\rightarrow 0$, by searches for a $5^{\text{th}}$ force, atomic parity violation and other experiments~\cite{{Bhattacharyya:2007pi},{Freitas:2007ip},{Liao:2007ic}}.} 
\label{direct}
\end{figure}
\par
By simple dimensional analysis of our expressions in Eqs.\,(\ref{scalarpi}) and (\ref{pscalarpi}), using Eq.\,(\ref{pot3FT}), we may expect the Unparticle corrections to the electromagnetic potential in the limit $\mu\rightarrow 0$ to be of a functional form
\be
V^{\text{\tiny{point}}}_{\mathcal{U}}(r, j)\sim~\frac{1}{r^{2j+1}}
\label{pointnunpot}
\eq
for $j\neq 1$. From Eq.\,(\ref{ordpf}) for the ordinary scalar/pseudo-scalar case, in the limit $m_{\Phi}\rightarrow 0$, we infer similarly that the potential may go dimensionally as $1/r^{3}$. This seems to suggest that, due to the generally non-integral values of the scaling dimension $j$, the fermiophobic Unparticle induced potential may have rich functional dependences. We will have more to say on this issue when we discuss the $j\rightarrow\, 1$ limit for Unparticle operators in the context of atomic energy-shifts.
\par
We can compare this fermiophobic scenario (where the correction to the potential is through photon vacuum polarizations) to the case of direct Unparticle scalar exchange between charged/uncharged fermions ( such a coupling could be generated radiatively at a higher order from fermiophobic couplings in the case of the charged particles) as shown in Fig.\,\ref{direct}. In the direct exchange case (fermiophilic)
\ba
U^{\text{\tiny{point}}}_{\mathcal{U}}(r, j)\sim~\int~\frac{d^{3}\vec{q}}{(2\pi)^{4}}~e^{i\vec{q}\cdot \vec{r}}~(q^{2})^{j-2}
\ea
 Therefore in this scenario we expect the correction to the potential to go like
\be
U^{\text{\tiny{point}}}_{\mathcal{U}}(r, j)\sim~\frac{1}{r^{2j-1}}
\eq
\par
Of course in the case of broken scale invariance the direct exchange will be Yukawa suppressed (by the scale breaking parameter $\mu$) leading to a finite range for the force. We are merely interested in the behavior of the direct-exchange potential as a comparison to the vacuum polarization case. As discussed in section I we will require $1\leq j< 2$ to be consistent with the Mack's unitarity condition and the requirement of UV insensitivity. This immediately implies that for the fermiophobic case under the point nucleus approximation the potential is always \textit{more singular} than $1/r^{2}$. This has some implications. We will look more closely at the \textit{sign} of the Unparticle Uehling corrections later, but for now let us assume that as $r\rightarrow 0$ the potential energy is negative. Consider an $\epsilon$-ball, near $r=0$, of radius $r_{\epsilon}$. Then from the uncertainty principle the mean energy at $r_{\epsilon}$ is
\ba
\langle E \rangle\sim~\frac{\hslash^{2}}{m\,r^{2}_{\epsilon}}-\frac{\xi^{2}}{r^{2j+1}_{\epsilon}}
\ea
where $\xi^2$ is a positive constant that is determined from some explicit computation. We have $(2j+1) > \,2\,\forall\, j\geq 1$ and this implies that for arbitrarily small $r_{\epsilon}$, the mean energy $\langle E \rangle$ is arbitrarily negative for a negative potential energy. Therefore the particle must fall to $r=0$ signifying an instability (see for example \cite{landau}). This may also be reasoned semi-classically from the criterion that the negative potential energy must be less singular than the centrifugal term, $l(l+1)/r^{2}$ in the Hamiltonian, to have a stable bound state. Note that for a positive potential energy ($\xi^{2}<\,0$) this constraint does not apply.  Also note that if we had not assumed the scale invariant Unparticle sector to be also conformally invariant then $j\geq\,1$ is no longer necessary, nevertheless for a negative potential energy  the scenario is still problematic for $\forall\,j>\,1/2$. In this case even for $j=\,1/2$ we have to require the condition
\ba
\xi^{2}~<~\frac{\hslash^{2}}{8\,m}
\ea
to get a stable bound state as is familiar from quantum mechanics~\cite{landau}. 
\par
Thus it seems, at least preliminarily, that the correction to the electromagnetic potential in the direct exchange case is in general less singular than the fermiophobic case for any given value of the scaling dimension ($j$) under the point nucleus approximation. These expectations are modified to a large extent by some physical considerations. The first obvious modification comes from the fact that near $r=0$ the finite size of the atomic nucleus becomes important and we would have to calculate the correction to the electromagnetic potential with an appropriate nuclear density profile. This would soften the singular nature of the potential near $r=0$. One may consider the expression in Eq.\,(\ref{pointnunpot}) as being roughly valid far away from the atomic nucleus in which case it may be approximated by a point source.  
\par
More importantly one needs to be careful in the evaluation to get consistent energy shifts as $j\rightarrow\,1^{+}$. From Eqs.\,(\ref{scalarpi}) and (\ref{pscalarpi}) we see that the $j\rightarrow\,1^{+}$ limit is not continuous. It seems to imply that if one calculates a potential and an energy shift using the quoted Unparticle polarization functions the energy shift increases without bound as one approaches $j=1$. This does not seem physically sensible. In the $j\rightarrow\,1^{+}$ limit, maintaining continuity throughout, one must expect to recover the ordinary scalar ($\phi$) and pseudo-scalar ($\tilde{\phi}$) situations. This feature seems to be peculiar to the Unparticle vacuum polarization diagram we are considering and is related to the fact that the momentum integrals diverge as $\Lambda^{2j}$, where $\Lambda$ is some momentum cut-off, and are only quadratically divergent when $j=1$.  In a similar calculation for the fermiophilic \textit{vector} Unparticle contributing to muon $(g-2)$ (where the photon in the loop is now replaced with a vector Unparticle) the relevant momentum integral, for $\mu\rightarrow 0$, is~\cite{Cheung:2007zza}
\ba
\frac{m_{\mu^{-}}^{2}}{\Lambda^{2j-2}_{\psi}}\int\,\frac{d^{4}l}{(2\pi)^{4}}\,\frac{1}{\left[l^{2}-\Delta_{\text{\tiny{V}}}(q^2,\,m^{2}_{\mu^{-}})+i\epsilon\right]^{4-j}}~\sim~\frac{m_{\mu^{-}}^{2}\,\Delta_{\text{\tiny{V}}}^{j-2}(q^2,\,m^{2}_{\mu^{-}})}{16\pi^2\,\Lambda^{2j-2}_{\psi}}
\ea
$m_{\mu^{-}}$ is the muon mass and $\Lambda_{\psi}$ is the interaction scale for the fermiophilic coupling. The limit $j\rightarrow1^{+}$ is now continuous and one obtains 
\ba
\frac{m_{\mu^{-}}^{2}\,\Delta_{\text{\tiny{V}}}^{j-2}(q^2,\,m^{2}_{\mu^{-}})}{\Lambda^{2j-2}_{\psi}}~\xrightarrow{j\rightarrow 1^{+}}~\frac{m_{\mu^{-}}^{2}}{16\pi^2\Delta_{\text{\tiny{V}}}(q^2,\,m^{2}_{\mu^{-}})}
\ea
as expected for an ordinary vector particle in the loop (like the photon).
\par
So to make the expressions (\ref{scalarpi}) and (\ref{pscalarpi})  consistent with the $j\rightarrow\,1^{+}$ limit and maintain continuity we will consider
\be
\frac{\Delta^{j-1}(q^2,\mu^2)}{j-1}~\xrightarrow{\forall j}~\left(M^{2}\right)^{j-1}\log\left[\frac{\Delta(q^2,\mu^2)}{M^{2}}\right]
\label{prescp}
\eq 
where a term that would lead to an infinite energy shift in the $j\rightarrow 1^{+}$ limit has been subtracted out. Here $M$ is again an arbitrary scale. As we mentioned previously for the ordinary scalar/pseudo-scalar case a natural and appropriate choice is to take $M\sim\Lambda_{\mathcal{U}}$, the scale of the infra-red fixed point, where the theory becomes scale-invariant and the description is in terms of Unparticles. The crude \textit{prescription} in Eq.\,(\ref{prescp}) is the one that we will follow in the present study to estimate Unparticle Uehling shifts for values of $j$ away from 1 and make the correspondence with the $j=1$ case. Note that for $j=1$ the prescription gives the ordinary scalar/pseudo-scalar expression. We will be interested mostly in the case when $j$ is close to $1$ where, as we shall see in the next section, the Unparticle induced atomic energy shifts have their maximum values. Also, as mentioned in section I, very close to $j=1$ the SM-SM contact terms are relatively less important numerically compared to the case when $j$ is very far from unity~\cite{Grinstein:2008qk}. We may therefore be optimistic that the above prescription will certainly capture the essential features of the Unparticle Uehling shifts in the interesting region near $j=1$.
\par
In the $\mu\rightarrow 0$ limit we may factor out the $q$ dependent terms in Eqs.\,(\ref{scalarpi}) and (\ref{pscalarpi}). Performing the integration over the Feynman variables we may show that \textit{the polarization functions for the scalar and pseudo-scalar Unparticles are exactly equal }$\forall\,j$:
\ba
\Pi_{\mathcal{O}}(-\vec{q}^{\,2},0,j)=\Pi_{\widetilde{\mathcal{O}}}(-\vec{q}^{\,2},0,j)
\ea
assuming $b=c$ and $\Lambda_{\gamma}=\tilde{\Lambda}_{\gamma}$. For the ordinary scalar/pseudo-scalar cases this equality may be understood readily in terms of the optical theorem. One consequence of our approximation in Eq.\,(\ref{prescp}) is that this relation is no longer exact when $j$ is sufficiently far from unity. We will therefore retain both the scalar and pseudo-scalar expressions in all our analyses.
\par
Let us now perform the analysis incorporating the finite extent of the nucleus. Assuming a static situation we have 
\ba
\delta A_{0}(\vec{r})=\int~\frac{d^{3}\vec{q}}{(2\pi)^{3}}~e^{i\vec{q}\cdot \vec{r}}~\hat{\Pi}_{\mathcal{U}}(-\vec{q}^{\,2}, \mu^2, j)~
\mathcal{G}^{\text{\tiny{photon}}}_{0\mu}(-\vec{q}^{\,2})~\mathcal{J}^{\mu}_{\text{\tiny{source}}}(\vec{q})
\label{potfinnu}
\ea
Let
\ba
\mathcal{J}^{0}_{\text{\tiny{source}}}(\vec{r})=\rho(\vec{r})=\,-Ze\,f(r)
\ea
where $f(r)$ is a suitable function, which we assume to be spherically symmetric for simplicity, describing the nuclear charge density profile. It must be suitably normalized such that 
\ba
\int~d^{3}\vec{r}~f(r)=~1
\ea
With this choice one may now perform the integration over the angular variables to obtain
\be
\delta A_{0}(\vec{r})=V_{\mathcal{U}}(r)\simeq~-Ze\int~\frac{d\lvert\vec{q}\rvert}{2\pi^{2}}~\frac{\sin(\lvert\vec{q}\rvert \,r)}{(\lvert\vec{q}\rvert\,r)}~\hat{\Pi}_{\mathcal{U}}(-\vec{q}^{\,2}, \mu^2, j)~\tilde{f}(\lvert\vec{q}\rvert)
\label{finuclpot}
\eq
where
\be
\tilde{f}(\lvert\vec{q}\rvert)=\int~d^{3}\vec{r}~e^{i\vec{q}\cdot \vec{r}}~f(r)
\eq
\par
We will choose a simple gaussian nuclear charge density profile
\be
f(r)=~f_{0}~e^{-r^{2}/(2\zeta)}
\label{nucp}
\eq
with
\ba
f_{0}=\frac{1}{(2\pi\zeta)^{3/2}}
\ea
to give the correct normalization. Note that at $r=\sqrt{2\zeta}$ the nuclear charge density is $1/e$ times its value at the origin. The gaussian nuclear profile may be considered to be an approximation to a more realistic two-parameter Fermi distribution. Usually the gaussian density profile is considered more appropriate for light nuclei~\cite{hnizdo}. In the next section we will see that we are interested in studying muonic lead, which is not a light nucleus. Nevertheless even with the simplistic choice of a gaussian profile we may expect the analysis to capture all the main features of the finite lead nucleus. With this \textit{ansatz}, for the nuclear profile, we expect any deviation from the actual case to be of $\mathcal{O}(1)$ and hence comparable to our ignorance about the factors $c,\,b$ in the effective Lagrangian. The main reason for choosing a gaussian profile is to simplify the subsequent analysis. The choice would also help us glean simpler analytic results which otherwise would have been more intractable and subject to numerical methods solely. 
\par
As previously mentioned for now we will assume exact scale invariance and set $\mu\rightarrow 0$. We will comment on this aspect later in section III and will incorporate a non-zero $\mu$ at that time. It will be seen that the main conclusions are not drastically changed for the case of broken scale invariance. Defining the dimensionless variable $z=\lvert\vec{q}\rvert\, r$ we may write Eq.\,(\ref{finuclpot}) for the two cases, using Eqs.\,(\ref{scalarpi}) and (\ref{pscalarpi}), as
\bea
\label{potscalar}
V_{\mathcal{O}}(r,j)&=&\frac{Z\, e\, c^2\,A_{j}}{64\pi^{4}\Lambda_{\gamma}^{2j}\sin(j\pi)}\int_{0}^{1} dx\,dy\,\delta(x+y-1)y^{1-j}(1-x)\left[(1+j)x-j\right]\\ \nn
&&\frac{(M^{2})^{j-1}}{j\,r^{3}}~\int_{0}^{\infty}dz~z~ \sin z~ e^{-\zeta z^{2}/(2 r^{2})}\left[\log(x(1-x))+\log\frac{z^{2}}{M^{2} r^{2}}\right]
\\
\label{potpscalar}
V_{\widetilde{\mathcal{O}}}(r,j)&=&-\frac{Z\, e\, b^2\,A_{j}}{64\pi^{4}\tilde{\Lambda}_{\gamma}^{2j}\sin(j\pi)}\int_{0}^{1} dx\,dy\, \delta(x+y-1)\left[ x(1-x)y^{1-j}\right]\\ \nn
&&\frac{(M^{2})^{j-1}}{j\,r^{3}}\int_{0}^{\infty}dz~z~\sin z~e^{-\zeta z^{2}/(2 r^{2})}\left[\log(x(1-x))+\log\frac{z^{2}}{M^{2} r^{2}}\right]
\eea
One may now perform the integration over the dimensionless parameter $z$ and the Feynman variables. The result may be expressed in terms of the exponential function and the generalized hypergeometric function $F^{p}_{q}[(a_{1},..,a_{p}),(b_{1},..,b_{q});w]$ in a compact form
\bea
\label{potillus}
V_{\mathcal{O}}(r,j)&=&~-\frac{Z\,e\,c^2}{64\,\pi^4}\left[B_{1}(j)\,\frac{e^{-\frac{r^2}{2 \zeta }} }{\zeta ^{3/2}}+~B_{2}(j)\,\frac{r^{2}\,e^{-\frac{r^2}{2 \zeta }} }{\zeta ^{5/2}}\ghf\left[(1, 1), (2, 5/2); r^2/(2 \zeta)\right]\right]\\\nn
V_{\widetilde{\mathcal{O}}}(r,j)&=&~-\frac{Z\,e\,b^2}{64\,\pi^4}\left[\tilde{B}_{1}(j)\,\frac{e^{-\frac{r^2}{2 \zeta }} }{\zeta ^{3/2}}+~\tilde{B}_{2}(j)\,\frac{r^{2}\,e^{-\frac{r^2}{2 \zeta }} }{\zeta ^{5/2}}\ghf\left[(1, 1), (2, 5/2); r^2/(2 \zeta)\right]\right]\\\nn
\eea
$B_{1}(j)$, $B_{2}(j)$, $\tilde{B}_{1}(j)$ and $\tilde{B}_{2}(j)$ are functions of the scaling dimension $j$ and are given by
\bea
\label{bl}
B_{1}(j)&=&-\frac{\sqrt{\pi}A_{j}\left(M^{2}\right)^{j-1}}{\sqrt{2}\Lambda^{2j}_{\gamma}\sin j \pi}\Bigg[\frac{2+j(j-2)(2 j-7)-(j-4) (j-3) (1+j(j-3))\har[3-j]}{j~(j-4)^2~ (j-3)^2}\\ \nn
&&~~~~~~~~~~~~~~~~~~~~~-\bigg\{\frac{j^2-3j+1}{j\,(j-4)\,(j-3)}\bigg\}\log\left(2M^2\zeta e^{\gamma_{E}-2}\right)\Bigg]\\ \nn
B_{2}(j)&=&-\frac{\sqrt{\pi}A_{j}\,\left(M^{2}\right)^{j-1}}{3\,\sqrt{2}\,\Lambda^{2j}_{\gamma}\,\sin j \pi}\left[\frac{-j^2+3j-1}{j\,(j-4)\,(j-3)}\right]\\ 
\label{btl}
\tilde{B}_{1}(j)&=&-\frac{\sqrt{\pi}A_{j}\left(M^{2}\right)^{j-1}}{\sqrt{2}\tilde{\Lambda}^{2j}_{\gamma}\sin j \pi}\Bigg[\frac{-5+j(5-j)+(j-4)(j-3)\har[4-j]}{j~(j-4)^2 ~(j-3)^2 }+\bigg\{\frac{1}{j(j-4)(j-3)}\bigg\}\log\left(2M^2\zeta e^{\gamma_{E}-2}\right)\Bigg]\\ \nn
\tilde{B}_{2}(j)&=&-\frac{\sqrt{\pi}A_{j}\,\left(M^{2}\right)^{j-1}}{3\,\sqrt{2}\,\tilde{\Lambda}^{2j}_{\gamma}\,\sin j \pi}\left[\frac{1}{j~(j-4)~(j-3)}\right]\\ \nn
\eea
where $\gamma_{E}$ is the Euler-Mascheroni constant and $\har[n]$ is the $n^{\text{\tiny{th}}}$ harmonic number given by $\sum_{k=1}^{n}(1/k)$ for integral values of $n$ and by the Euler integral representation
\ba
\har[n]=\int_{0}^{1}dx~\frac{1-x^{n}}{1-x}
\ea
for $n\notin\,+\mathbb{Z}$. Note that $\sqrt{\zeta}$ has dimensions of length and hence the above expressions for the potentials have the expected dimensions. 
\begin{figure}
\includegraphics[width=12cm,angle=0]{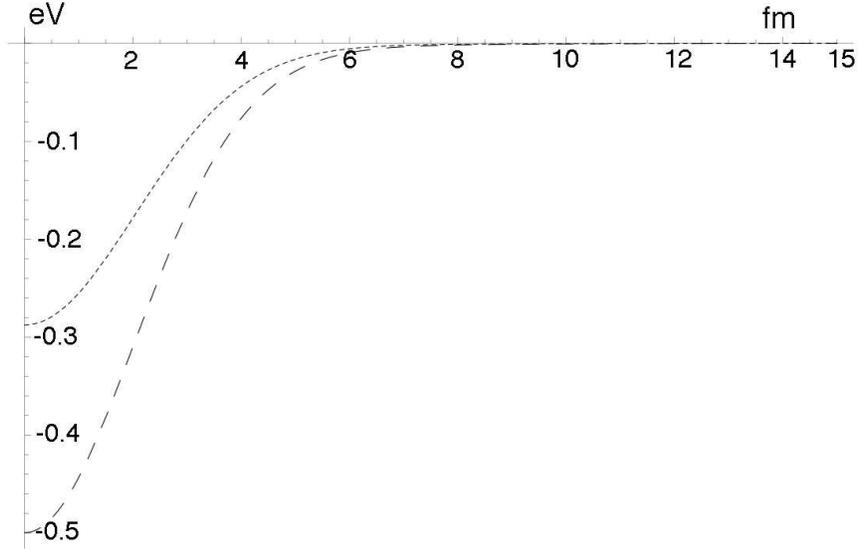}
\caption{Potential energies in muonic lead for the pseudo-scalar Unparticle case based on Eq.\,(\ref{potpscalar}), assuming perfect scale invariance $\mu\rightarrow 0$, for $j=1.01$ (dashed line) and $1.15$ (small-dashed line). We have taken $b\simeq\mathcal{O}(1)$, $\zeta\simeq\,4\,\text{fm}^{2}$ (to model the case of a $\mu^{-}-\text{Pb}^{\,208}_{\,82}$ nucleus) and adopted a reference value for the interaction scale $\tilde{\Lambda}_{\gamma}\sim\,246\,\text{GeV}$. Note that the potential goes to zero very rapidly away from $r=0$ and has a constant value at the origin implying that there is no singular behavior. }
\label{potPS}
\end{figure}
The functional forms of the potential energies in the pseudo-scalar Unparticle case, $e\,V_{\widetilde{\mathcal{O}}}(r,j)$, for two different values of the scaling dimension $j$ are shown in Fig.\,\ref{potPS}. The profile indicates that for the \textit{pseudo-scalar} Unparticle the potential energy is \textit{negative}. It is seen that the singular nature of the Unparticle Uehling correction, near $r=0$, has been eliminated by taking into account the finite size of the nucleus. We will explore in more detail the sign of the Unparticle potential energies shortly.
 \par
 The Unparticle Uehling potential would cause a small energy shift in the atomic energy levels. The energy level structure in a bound system may be deduced (without explicitly calculating all the energies) by studying the relevant potential~\cite{{Quigg:1979vr}}. We will apply some of these methods to study the level structure of the \textit{energy-shifts} ($\delta E^{\mathcal{U}}$) due to Unparticle Uehling potentials (\ref{potscalar}) and (\ref{potpscalar}).
 \par
 A general result~\cite{{Grosse:1983ii}} is that for a potential $V(r)$ depending on 
 \ba
 e\,\nabla^{2}\,V(r)\equiv~\frac{e}{r^{2}}\frac{d}{dr}\left(r^{2}\,\frac{dV(r)}{dr}\right)\gtreqqless~0
 \ea
 the energy levels are ordered as
 \ba
 E(n,\,l)~\gtreqqless~E(n,\,l+1)
 \ea
 Here $n$ is the principal quantum number and $l$ is the angular momentum quantum number. Note that for the simple Coulomb potential $\nabla^{2}\,V(r)=0$ for non-zero $r$ and gives the familiar result that energy levels depend only on the principal quantum number $n$. For the Unparticle Uehling potentials being considered we see that the situation may be complicated somewhat by the fact that for the finite nucleus the Laplacian of the potential energy may change sign. Thus strictly speaking we must consider if
 \be
 \label{lapavg}
 \Big\langle\, e\,\nabla^{2}\,V_{\mathcal{U}}(r,\,j,\,\mu^{2} )\,\Big\rangle_{n,\,l+1}~ \gtreqqless~0
 \eq
 Here the averaging $\langle\ldots\rangle_{n,\,l+1}$ is over the relevant $(n,\,l+1)$ wavefunction~\cite{Grosse:1983ii}. Depending on the sign of the above expression we would have a relation between the Unparticle induced energy shifts
 \ba
 \delta E^{\mathcal{U}}(n,\,l,\,j,\,\mu^{2})~\gtreqqless~\delta E^{\mathcal{U}}(n,\,l+1,\,j,\, \mu^{2})
 \ea
 \par
 Let us proceed to study the Laplacian of the potentials in Eqs.\,(\ref{potscalar}) and (\ref{potpscalar}). As mentioned before we will consider the case of $\mu\simeq\,0$ presently and concentrate on the low-lying states, i.e. states near $r=0$. It is noted, from Eqs.\,(\ref{bl}) and (\ref{btl}), that for both the scalar and pseudo-scalar Unparticle potentials
 \ba
 B_{1}(j),~B_{2}(j)~>~0\\
 \tilde{B}_{1}(j),~\tilde{B}_{2}(j)~>~0\\
 \ea
 for $\forall j\in[1,2)$. It is also seen that
 \ba
  B_{1}(j)~\gg~B_{2}(j)  \\ 
   \tilde{B}_{1}(j)~\gg~ \tilde{B}_{2}(j)  \\ 
 \ea
  Thus we infer that both \textit{pseudo-scalar and scalar Unparticle Uehling potential energies are negative $\forall\,j\,\in[1,2)$}. This reaffirms the indications from Fig.\,\ref{potPS}. It also means, as indicated by the $B_{1}(j)$ and $\tilde{B}_{1}(j)$ coefficients, that the first term in the Unparticle Uehling potential dominates over the second term. The Laplacians of the Uehling potentials in Eq.\,(\ref{potillus}) are given by
\bea
\label{lappot}
 e\,\nabla^{2}\, V_{\mathcal{O}}(r,j,0)&=&-\frac{Z\,e^2\,c^2}{64\,\pi^4}\Bigg[ B_{1}(j)\,\frac{e^{-\frac{r^2}{2 \zeta }}}{\zeta^{7/2}}\,\left(r^{2}-3\zeta\right)+ B_{2}(j)
 \frac{e^{-\frac{r^2}{2 \zeta }} }{r \zeta ^{9/2}}\bigg\{12r\zeta ^2-3 e^{\frac{r^2}{2 \zeta }} \sqrt{2 \pi } \zeta ^{5/2} \erf\left[\frac{r}{\sqrt{2\zeta}}\right]\\ \nn
&& ~~~~~~~~~~~~~+r^3 \left(r^2-3 \zeta \right) \ghf\left[\{1,1\},(2,5/2); r^2/(2 \zeta)\right]\bigg\}\Bigg]\\ \nn
e\,\nabla^{2}\,V_{\widetilde{\mathcal{O}}}(r,j,0)~&=&-\frac{Z\,e^2\,b^2}{64\,\pi^4}\Bigg[\tilde{B}_{1}(j)\,\frac{e^{-\frac{r^2}{2 \zeta }}}{\zeta^{7/2}}\,\left(r^{2}-3\zeta\right)+\,\tilde{B}_{2}(j)\frac{e^{-\frac{r^2}{2 \zeta }} }{r \zeta ^{9/2}}\bigg\{12r\zeta ^2-3 e^{\frac{r^2}{2 \zeta }} \sqrt{2 \pi } \zeta ^{5/2} \erf\left[\frac{r}{\sqrt{2\zeta}}\right]\\ \nn
&& ~~~~~~~~~~~~~+r^3 \left(r^2-3 \zeta \right) \ghf\left[\{1,1\},(2,5/2); r^2/(2 \zeta)\right]\bigg\}\Bigg]
\eea
where $\erf[z]$ is the error function defined as
\ba
\erf[z]=\frac{2}{\sqrt{\pi}}\int_{0}^{z}dt~e^{-t^{2}}
\ea
The Laplacians of the Unparticle Uehling potentials correspond to vacuum charge densities created by the Unparticles
\ba
\rho^{\text{\tiny{VP}}}_{\mathcal{O}}(r,j)=~-\,\epsilon_{0}\,\nabla^{2}V_{\mathcal{O}}(r,j,0)\\ \nn
\rho^{\text{\tiny{VP}}}_{\widetilde{\mathcal{O}}}(r,j)=~-\,\epsilon_{0}\,\nabla^{2}V_{\widetilde{\mathcal{O}}}(r,j,0)
\ea
Since the potentials in Eq.\,(\ref{potillus}) fall off faster than $1/r$ (as $r\rightarrow\infty$) the net, Unparticle induced, vacuum charge vanishes trivially
\ba
Q^{\text{\tiny{VP}}}_{\mathcal{U}}=\int d^{3}r\,\rho^{\text{\tiny{VP}}}_{\mathcal{U}}(r,j)=\int d^{3}r\left[-\epsilon_{0}\nabla^{2}V_{\mathcal{U}}(r,j,0)\right]=\,-\epsilon_{0}\,\lim_{R\rightarrow\infty} \oint_{R} d\vec{\sigma}\cdot\vec{\nabla}V_{\mathcal{U}}(r,j,0)=~0
\ea
as we should expect.
\par
Now from our analytical expressions it is observed, as mentioned above, that for typical values of the scaling dimension $j$ and radial coordinate $r$, the first term in the Unparticle Uehling potential dominates. It is seen from Eq.\,(\ref{lappot}) that the first term changes sign at $r=\sqrt{3\zeta}$. Thus we may suspect that the sign of the Laplacian for the complete Uehling potential would also be dominated by the sign of the first term. Therefore we have for both the potentials
\bea
e\,\nabla^{2}\, V_{\mathcal{O}}(r,j,0) ~\gtrless~0~~~\forall~r~\lessgtr~\sqrt{3\zeta}\\ \nn
e\,\nabla^{2}\, V_{\widetilde{\mathcal{O}}}(r,j,0) ~\gtrless~0~~~\forall~r~\lessgtr~\sqrt{3\zeta}
\eea
These expectations are confirmed by explicit numerical computations where it is found that the Laplacian of the Unparticle Uehling potential indeed changes sign at $\sqrt{3\zeta}$.
\par
We are interested in the low-lying atomic states which will have relatively large energy shifts due to the Unparticle Uehling corrections. The higher angular momentum ($p,\,d,\ldots$) wavefunctions are spread out from the origin. Specifically, it is seen that they are mainly non-zero for $r>\sqrt{3\zeta}$ for semi-realistic values of $\zeta$ that will model an atomic nucleus ( for example $\zeta\simeq\,4\,\text{fm}^{2}$ for the lead nucleus). From Eq.\,(\ref{lapavg}) we therefore note that due to this, the sign of $\langle\, \nabla^{2}\,V_{\mathcal{U}}(r,\,j,\,\mu^{2} )\rangle_{n,\,l+1}$ will be dominated by the sign of the Laplacian in the region $r>\sqrt{3\zeta}$. Thus for the Unparticle scalar and pseudo-scalar Uehling potentials
\bea
 \label{lapavgSPS}
 \Big\langle\, e\,\nabla^{2}\,V_{\mathcal{O}}(r,\,j,\,\mu^{2} )\,\Big\rangle_{n,\,l+1}~<~0\\ \nn
 \Big\langle\, e\,\nabla^{2}\,V_{\widetilde{\mathcal{O}}}(r,\,j,\,\mu^{2} )\,\Big\rangle_{n,\,l+1}~<~0
 \eea
 This implies that in the case of perfect scale invariance we have 
\bea
\label{SPSenghier}
\delta E^{\mathcal{O}}(n,\,l,\,j,\,0)~&<&~\delta E^{\mathcal{O}}(n,\,l+1,\,j,\, 0)\\ \nn
\delta E^{\widetilde{\mathcal{O}}}(n,\,l,\,j,\,0)~&<&~\delta E^{\widetilde{\mathcal{O}}}(n,\,l+1,\,j,\, 0)
\eea
This would mean that if we were looking specifically at $l=0,\,1\,\text{and }2$ states near $r=0$ the Unparticle Uehling corrections would be ordered as 
\ba
\delta E^{\mathcal{O}}_{n,\,s}(j)~<~\delta E^{\mathcal{O}}_{n,\,p}(j)~<~\delta E^{\mathcal{O}}_{n,\,d}(j)\\ 
\delta E^{\widetilde{\mathcal{O}}}_{n,\,s}(j)~<~\delta E^{\widetilde{\mathcal{O}}}_{n,\,p}(j)~<~\delta E^{\mathcal{O}}_{n,\,d}(j)
\ea
We will explore the level structure of the Unparticle Uehling shifts in the next section and confirm explicitly the general results above.
\end{section}


\begin{section}{Muonic atoms and energy shifts}
\par
Muonic atoms are created when muons are stopped in matter~\cite{Wheeler:1949zz}. The muon initially undergoes scattering and as it loses energy it may be captured by one of the atoms in a higher orbital. From here it cascades down, via various processes, to the innermost orbits where it persists until decay. The whole cascade process is expected to occur within $\sim 10^{-9}-10^{-12}\,\text{s}$ and the muon spends $\sim 10^{-7}-10^{-6}\,\text{s}$ in the inner orbits until decay. During cascade and its time in the innermost orbits we may probe the energy levels of the system (see \cite{Borie:1982ax} and references therein for instance). 
\par
Muonic atoms are especially suited for probing the effects of oblique corrections to the photon propagator~\cite{{Wheeler:1949zz}, {Borie:1982ax}}. In the QED induced hydrogenic Lamb shift~\cite{lamb} the $2s_{1/2}$ state is found to lie \textit{above} the $2p_{1/2}$ state by
\ba
\delta E^{\text{\tiny{total}}}_{\text{\tiny{Lamb}}}=~E_{2s_{1/2}}-E_{2p_{1/2}}\simeq~1058~\text{MHz}
\ea
which corresponds to about $4.4\times 10^{-6}\,\text{eV}$. The QED vacuum polarization contribution to the Lamb shift ~\cite{Bethe:1947id} on the other hand is found to be
\ba
\delta E^{\text{\tiny{VP}}}_{\text{\tiny{Lamb}}}=~E_{2s_{1/2}}-E_{2p_{1/2}}\simeq~-27.1~\text{MHz}
\ea
For later comparison to the Unparticle case we note that the above $e^{+}-e^{-}$ vacuum polarization contribution is about $1\times 10^{-7}\,\text{eV}$. Thus in ordinary atoms the vacuum polarization effects (oblique corrections) contribute less significantly than the other QED corrections. Things are dramatically different in the case of muonic atoms owing to the much higher mass of the muon compared to the electron. Now, the energy shift due to vacuum polarization may be related to the Uehling correction $\delta V(\vec{r})$ as
\be
\delta E^{\text{\tiny{VP}}}_{\,nl}=~\int~d^{3}r~\Psi^{*}_{nl}(\vec{r})~ e\,\delta V(\vec{r})~\Psi_{nl}(\vec{r})
\label{eshiftexp}
\eq
We are specifically interested in the low-lying atomic states and the $S$ and $P$ states are particularly attractive since the modification to the potential, as we saw previously, mostly occurs near $r=0$ and falls off sharply as one goes away from the center. Taking the wavefunction $\Psi_{nl}(\vec{r})$, to lowest order, to be the Schr\"odinger wavefunction we note that near $r=0$
\ba
\lvert\Psi^{e^{-}}_{nl}\rvert^{2}\sim~m^{3}_{e^{-}}
\ea
Thus we expect that in muonic atoms, where an electron in the inner orbital is replaced by a muon, the energy shift due to oblique corrections would be enhanced as
\ba
\delta E^{\mu^{-}}_{\,nl}\simeq~\left(\frac{m_{\mu^{-}}}{m_{e^{-}}}\right)^{3}~\delta E^{e^{-}}_{\,nl}
\ea
Putting the masses in, $\left(m_{\mu^{-}}/m_{e^{-}}\right)^{3} \simeq~10^{7}$. This is a huge enhancement to the vacuum polarization contribution. In fact it is found that in muonic hydrogen the  $2s_{1/2}$ state lies \textit{below} the $2p_{1/2}$ level with QED vacuum polarization contributing about $206\,\text{meV}$ compared to about $0.6\,\text{meV}$ from other QED corrections~\cite{Borie:1982ax}. Another way to understand this enhancement is by looking at the wavefunction profiles in muonic atoms and noting that the wavefunction penetrates into the nucleus much more than in the electron case. In all our numerical computations we use the hydrogenic Schr\"odinger wavefunctions to lowest order, with $m_{e^{-}}$ replaced everywhere with $m_{\mu^{-}}$ and taking the appropriate value for the atomic number $Z$. One must in principle use the bound state solutions to the Dirac equation but for the purposes of our estimations the Schr\"odinger wavefunctions are sufficent. One aspect that we would like to point out though is that the bound state solutions to the Dirac equation are more localized near $r=0$ compared to the Schr\"odinger wavefunctions and hence will in general yield a slightly higher estimate for the corrections to the energy levels. Any error that comes from this simplification will be at most an $\mathcal{O}(1)$ factor and similar to our ignorance regarding the coefficients in the Lagrangian or the interaction scales. Thus all calculated energy shifts in our study must be interpreted as accurate only up to undetermined $\mathcal{O}(1)$ factors to accommodate our ignorance of the interaction scales, coefficients and other approximations we have made.
 \par
The typical binding energy of an atom with atomic number $Z$ goes as
\ba
E_{\text{\tiny{Bound}}}\sim~Z^{2}
\ea
But from Eq.\,(\ref{eshiftexp}) we see that
\ba
\delta E^{\text{\tiny{VP}}}_{nl}\sim~Z^{4}
\ea
since
\ba
\lvert\Psi_{nl}\rvert^{2}\sim~Z^{3}~,~~\delta V(\vec{r})\sim~Z
\ea
Based on the above expressions we expect 
\ba
\frac{\delta E^{\text{\tiny{VP}}}_{nl}}{E_{\text{\tiny{Bound}}}}\sim~Z^{2}
\ea	
This motivates the reason for choosing \textit{a high $Z$ system} while probing for the existence of tiny Unparticle induced energy shifts. 
\par
Based on the above considerations two very favorable systems for probing Unparticle vacuum polarization induced energy shifts are muonic mercury ($\mu^{-}-\text{Hg}^{\,200}_{\,80}$) and muonic lead ($\mu^{-}-\text{Pb}^{\,208}_{\,82}$). We will choose as our reference system muonic lead. With $Z_{\text{\tiny{Pb}}}=82$ and $A_{\text{\tiny{Pb}}}=208$, some of the relevant scales in the muonic lead system are
\ba
m_{\mu^{-}}\simeq~0.1\,\text{GeV}~,~~\lambda_{\text{\tiny{Compton}}}^{\mu^{-}}\simeq~12\,\text{fm}~,~~r_{\text{\tiny{Bohr}}}^{\mu^{-}}\simeq~3\,\text{fm}~,~~r_{\text{\tiny{Nucleus}}}^{\text{\tiny{Pb}}}\approx~R_{0}\,A_{\text{\tiny{Pb}}}^{1/3}\simeq~7\,\text{fm}
\ea
where $R_{0}\simeq~1.2\,\text{fm}$. From these crude estimates we see that the muonic Bohr radius in lead is much smaller than the extent of the lead nucleus or the muonic Compton wavelength and hence the muonic wavefunction will penetrate into the nucleus to a large extent compared to electronic wavefunctions. This results in an enhancement of the Uehling energy shift as we argued previously. It also implies that finite nuclear effects would be much more important in the low-lying states of muonic atoms, especially the muonic lead system we are considering~\cite{{Chen:1970bn}, {skar}}. 
\par
Another point to note is that for lead, with $Z=82$, the parameter $\alpha\,Z$ used in perturbative QED calculations is no longer small. This makes the QED calculations more complicated and leads to larger theoretical uncertainties. Table I shows some of the QED corrections to the low-lying states of muonic lead. The interested reader is referred to theoretical details in \cite{Borie:1982ax} and references therein. 
\par
\vspace{7mm}
TABLE I. QED corrections, in eV, for the low-lying states in $\mu^{-}-\text{Pb}^{\,208}_{\,82}$ (see~\cite{{Borie:1982ax},{Haga:2007mx}} and references therein).
\vspace{3mm}\\
\begin{tabular}{ccccccccc}
\hline
\hline
QED corrections &
\hspace{6mm}$1s_{1/2}$\hspace{6mm}&
\hspace{6mm}$2s_{1/2}$\hspace{6mm}&
\hspace{6mm}$2p_{1/2}$\hspace{6mm}&
\hspace{6mm}$2p_{3/2}$\hspace{6mm}&
\hspace{6mm}$3p_{1/2}$\hspace{6mm}&
\hspace{6mm}$3p_{3/2}$\hspace{6mm}&
\hspace{6mm}$3d_{3/2}$\hspace{6mm}&
\hspace{6mm}$3d_{5/2}$\hspace{6mm}\\
\hline
Electronic Uehling and K\"{a}ll\'{e}n-Sabry & -67864 & -19537 & -32648 & -30082 & -10871 & -10334 & -10605 & -9941 \\
Electronic Wichman-Kroll &  492 & 244 & 348 & 335 & 160 & 160  & 186 & 180  \\
Muonic Uehling corrections & -248  & -43 & -45 & -34 & -14 & -11  & -1 & -1  \\
Leading self energy corrections & 3220  & 696 & 348 & 649 & 149 & 224  & -44 & 51  \\
Higher order self energy corrections & 153  & 25 & 65 & 58 & 21 & 20  & 8 & 6  \\
Electron screening & -5  & -25 & -13 & -13 & -52 & -54  & -37 & -39  \\
Recoil correction& -382  & -87 & -111 & -95 & -30 & -26  & -15 & -14  \\
\hline\hline
\end{tabular}
\vspace{7mm}
\par
Similarly the various finite nuclear effects can be quantified to a large extent in muonic $\text{Pb}^{\,208}_{\,82}$ (see M.-y. Chen (1970)\,\cite{Chen:1970bn}, H. F. Skardhamar (1970)\,\cite{skar} and A. Haga \textit{et al.} (2007)\,\cite{Haga:2007mx}, for example). In Table II we give the current estimates of nuclear polarization (NP) effects in $\mu^{-}-\text{Pb}^{\,208}_{\,82}$ for three different models of transition densities as quoted in~\cite{Haga:2007mx}. 
\par
\vspace{7mm}
TABLE II. Estimates of the NP effects (eV) in $\mu^{-}-\text{Pb}^{\,208}_{\,82}$ taken from~\cite{Haga:2007mx}. The gauge dependences are shown in brackets.
\vspace{3mm}\\
\begin{tabular}{ccccccccc}
\hline
\hline
Nuclear polarization corrections &
\hspace{6mm}$1s_{1/2}$\hspace{6mm}&
\hspace{6mm}$2s_{1/2}$\hspace{6mm}&
\hspace{6mm}$2p_{1/2}$\hspace{6mm}&
\hspace{6mm}$2p_{3/2}$\hspace{6mm}&
\hspace{6mm}$3p_{1/2}$\hspace{6mm}&
\hspace{6mm}$3p_{3/2}$\hspace{6mm}&
\hspace{6mm}$3d_{3/2}$\hspace{6mm}&
\hspace{6mm}$3d_{5/2}$\hspace{6mm}\\
\hline
TGT model & -2727(4)& -463(1) & -1357(7) & -1425(9) & -561(4) & -749(1) & -226(0) & -43(0) \\
RIN model & -3599(10) & -611(4)& -1590(10) & -1656(10) & -690(3) & -914(1)  & -239(0) & -42(0)  \\
JS model & -5721(28) & -930(8) & -2178(13) & -2214(7) & -929(3) & -1179(2)  & -280(0) & -38(0)  \\
\hline\hline
\end{tabular}
\vspace{7mm}
\par
It was noted in the past that there is a discrepancy in the $\Delta 2p$ and $\Delta 3 p$ NP calculations with results from muonic lead spectroscopy~\cite{{Bergem:1988zz},{Kessler:1975ju}}. This discrepancy seems to have been partially tackled in the work of A. Haga \textit{et al.}~\cite{Haga:2007mx} and the current discrepancy for $\Delta 2p$ is in the ball-park of about 50 eV as shown in Table III.
\par
\vspace{7mm}
TABLE III. Comparison of $\Delta p$ splittings (KeV) in $\mu^{-}-\text{Pb}^{\,208}_{\,82}$ \cite{Haga:2007mx} based on the QED results in Table I and NP calculations in Table II.
\vspace{3mm}\\
\begin{tabular}{ccccccccc}
\hline
\hline
$\mu^{-}-\text{Pb}^{\,208}_{\,82}$ level &
\hspace{6mm}TGT\hspace{6mm}&
\hspace{6mm}RIN\hspace{6mm}&
\hspace{6mm}JS\hspace{6mm}&
\hspace{6mm}Exp.\hspace{6mm}&\\
\hline
$2p_{3/2}-2p_{1/2}$ & 184.858 & 184.846 & 184.829 & 184.788(27)  \\
$3p_{3/2}-3p_{1/2}$ &  47.231 & 47.208& 47.225 & 47.197(45)  \\
\hline\hline
\end{tabular}
\vspace{7mm}
\par
Historically the spectroscopic information from low-lying muonic lead transitions was used to constrain parameters in the theoretical nuclear calculations. In Table IV the results of precision measurements on some of the low-lying transitions in muonic lead are shown. It is noted that the experimental uncertainties are typically of the order of a few tens of eV.
\par
\vspace{7mm}
TABLE IV. Precision measurements on some of the low-lying $\mu^{-}-\text{Pb}^{\,208}_{\,82}$ transitions~\cite{{Bergem:1988zz},{Kessler:1975ju}}.
\vspace{3mm}\\
\begin{tabular}{ccc}
\hline
\hline
Muonic transition~~~&~~~Energy$^{a}$ (KeV)~~~&~~~Energy$^{b}$ (KeV)\\
$2p_{3/2}\leftrightarrow1s_{1/2}$ & 5962.770(420) & 5962.854(90)\\
$2p_{1/2}\leftrightarrow1s_{1/2}$ & 5777.910(400) & 5778.058(100)  \\
$3d_{3/2}\leftrightarrow2p_{1/2}$ & 2642.110(60)   & 2642.332(30) \\
$3d_{5/2}\leftrightarrow2p_{3/2}$ & 2500.330(60 ) & 2500.590(60) \\
$3d_{3/2}\leftrightarrow2p_{3/2}$ & 2457.200(200)  & 2457.569(70)   \\
$3p_{3/2}\leftrightarrow2s_{1/2}$ & 1507.480(260) & 1507.754(70)   \\
$3p_{1/2}\leftrightarrow2s_{1/2}$ & $-$ & 1460.558(32)   \\
$2s_{1/2}\leftrightarrow2p_{1/2}$ & 1215.430(260)  & 1215.330(30)   \\
$2s_{1/2}\leftrightarrow2p_{3/2}$ & 1030.440(170) & 1030.543(27)   \\
\hline\hline
\end{tabular}
\begin{description}
\item{$^{a}$}
D. Kessler \textit{et al.} (1975)
\item{$^{b}$}
P. Bergam \textit{et al.} (1988)
\end{description}
\vspace{7mm}
\par
As suggested by Fig.\,\ref{potPS} the states $1S$ and $2S$ are most affected by the Unparticle Uehling potential. A transition from $1S$ to $2S$ is nevertheless forbidden due to the electric dipole selection rule $\Delta l=\pm 1$. That leaves the possibility of a transition to an $l=1$ state, that might still be sensitive to the Unparticle Uehling effect. We will consider, as the prototypical low-lying muonic transition, excitations between $1S$ and $2P$. Based on a direct fit to muonic lead transition energy data (including higher level transitions), keeping the variables in the Fermi distribution as free parameters, one may obtain an estimate for the discrepancy between theory and experiment in the $1S-2P$ transition in $\mu^{-}-\text{Pb}^{\,208}_{\,82}$~\cite{Bergem:1988zz} as
\bea
\label{thexpdisc}
\Delta E_{\text{\tiny{Exp.}}}\left[1s_{1/2}-2p_{1/2}\right]-\,\Delta E_{\text{\tiny{Calc./Fit}}}\left[1s_{1/2}-2p_{1/2}\right]&\simeq&\,227\,\text{eV}\\ \nn
\Delta E_{\text{\tiny{Exp.}}}\left[1s_{1/2}-2p_{3/2}\right]-\,\Delta E_{\text{\tiny{Calc./Fit}}}\left[1s_{1/2}-2p_{3/2}\right]&\simeq&\,-89\,\text{eV}
\eea
This estimate is a very conservative one based on a direct fit~\cite{Bergem:1988zz} with a very poor $\chi^2/\text{d.o.f}=187$. We will take the above estimate as a crude measure of the discrepancy between measurement and calculation. It should be noted also that the NP values used in Eq. (\ref{thexpdisc}) are obtained keeping the variables in the Fermi distribution as free parameters and are therefore not completely theoretical. 
\par
We will see shortly that, for \textit{typical} values of the model parameters, this makes the identification of any possible Unparticle Uehling corrections in $\mu^{-}-\text{Pb}^{\,208}_{\,82}$ extremely difficult since the magnitude of any such correction will be found to be well below the NP uncertainties. This would imply that any Unparticle Uehling correction if present may not probably be unambiguously seen due to the data-fitting procedure required to extract information for the theoretical nuclear calculation in $\mu^{-}-\text{Pb}^{\,208}_{\,82}$, unless we can obtain such information independently. 
\par
There are two scenarios where these conclusions could get modified. The first, as we commented earlier in section I, is if the UV sector has a very large fermion \textit{multiplicity} (large $N_{f}$) which increases the total contribution from the fermiophobic Unparticle sector. The second is muonic atoms with intermediate-Z which may be more amenable to nuclear polarization calculations.

We now proceed to analyze the $1S-2P$ level corrections in various cases of interest.
\par
Let us first consider the case of ordinary scalars ($\phi$) and pseudo-scalars ($\tilde{\phi}$). The relevant corrections to the electromagnetic potential is obtained from expressions in Eq.\,(\ref{ordpf}). In this case we find that the energy shift due to scalar oblique corrections is
\bea
\label{eshiftscalar}
\delta E_{\,nl}^{\phi}&=&~\int~d^{3}r~\lvert\Psi^{\mu^{-}}_{nl}(\vec{r})\rvert^{2}~ e\,\delta V_{\phi}(\vec{r}) \\ \nn
&=&-\frac{Z_{\text{\tiny{Pb}}}\,e^2\,a^2}{16\pi^2\Lambda_{\phi}^{2}}\int~d^{3}r~\lvert\Psi^{\mu^{-}}_{nl}(\vec{r})\rvert^{2}\int~\frac{d^{3}\vec{q}}{(2\pi)^{3}}~e^{i\vec{q}\cdot \vec{r}}(\vec{q}^{\,2})^{-1}\tilde{f}(\vec{q})\int_{0}^{1}~dx\,\bigg\{\Delta^{'}(q^{2},m_{\phi}^{2})-\vec{q}^{\,2}(x-1)^{2}\bigg\}\,\log\frac{\Delta^{'}(q^{2},m_{\phi}^{2})}{M^{2}}
\eea
where 
\ba
\Delta^{'}(q^2,m_{\phi}^2)=\,x(x-1)\,q^{2}+\,(1-x) \,m_{\phi}^{2}
\ea
and $m_{\phi}$ is the mass of the scalar as before. 
\par
For the pseudo-scalar case similarly
\bea
\label{eshiftpscalar}
\delta E_{\,nl}^{\tilde{\phi}}&=&+\frac{Z_{\text{\tiny{Pb}}}\,e^2\,\tilde{a}^2}{16\pi^2\Lambda_{\tilde{\phi}}^{2}}\int~d^{3}r~\lvert\Psi^{\mu^{-}}_{nl}(\vec{r})\rvert^{2}\int~\frac{d^{3}\vec{q}}{(2\pi)^{3}}~e^{i\vec{q}\cdot \vec{r}}(\vec{q}^{\,2})^{-1}\tilde{f}(\vec{q})\int_{0}^{1}~dx\,\Delta^{''}(q^{2},m_{\tilde{\phi}}^{2})\,\log\frac{\Delta^{''}(q^{2},m_{\tilde{\phi}}^{2})}{M^{2}}
\eea
and
\ba
\Delta^{''}(q^2,m_{\phi}^2)=\,x(x-1)\,q^{2}+\,(1-x) \,m_{\tilde{\phi}}^{2}
\ea
$m_{\tilde{\phi}}$ denotes the mass of the pseudo-scalar. Let us consider the case when $m_{\phi}=0$ and $m_{\tilde{\phi}}=0$. With this choice the above expressions may be evaluated numerically to calculate the energy shift in the muonic lead $1S-2P$ transition. We take $a,\,\tilde{a}\simeq\,2$  and adopt the reference value $\Lambda_{\Phi}\sim246\,\text{GeV}$ with a renormalization scale $M\simeq\Lambda_{\Phi}$. With these parameters the magnitude of the energy shift in the $1S-2P$ transition for $m_{\Phi}=0$ is estimated to be
\be
\Delta E^{\Phi}_{2p-1s}=\lvert\delta E_{\,2p}^{\phi}-\delta E_{\,1s}^{\phi}\rvert\simeq~\lvert\delta E_{\,2p}^{\tilde{\phi}}-\delta E_{\,1s}^{\tilde{\phi}}\rvert\simeq~0.140\,\text{eV}\sim~\mathcal{O}(0.1)\,\text{eV}
\eq
It is found that the correction to the transition, from the SM value, is \textit{positive for both  scalars and pseudo-scalars}. The above estimate may be enhanced due to a large \textit{multiplicity} in the UV fermion sector and is not uncommon in many string-inspired QCD-like models~\cite{Sakai:2004cn} as we commented earlier. 
\par
 Let us consider the same calculation for an Unparticle scalar/pseudo-scalar of scaling dimension $j$. We parametrize $j$ as a deviation from the ordinary scalar/pseudo-scalar case by putting
\ba
j=1+\eta
\ea
As argued in section I  we require $0\leq\eta<1$. Then the energy shifts due to the scalar and pseudo-scalar Unparticle oblique corrections are respectively
\bea
\label{seshift}
\delta E_{\,nl}^{\mathcal{O}}(j)&=&\frac{Z_{\text{\tiny{Pb}}}\,e^2\,c^2A_{j}}{2\Lambda_{\gamma}^{2j}\sin(j\pi)}\int d^{3}r~\lvert\Psi^{\mu^{-}}_{nl}(\vec{r})\rvert^{2}\int\frac{d^{3}\vec{q}}{(2\pi)^{3}}~e^{i\vec{q}\cdot \vec{r}}(\vec{q}^{\,2})^{-1}\tilde{f}(\vec{q})\int dx\,dy\,\delta(x+y-1)\,y^{1-j}\\ \nn
&&\frac{\left(M^{2}\right)^{j-1}}{(4\pi)^{2}}\left[-\vec{q}^{\,2}(x-1)^{2}\log\Delta(q^{2},\mu^2)/M^{2}+\Delta(q^{2},\mu^2)~\frac{\log\Delta(q^{2},\mu^2)/M^{2}}{\left(1+\eta\right)}\right]
\eea
\bea
\label{pseshift}
\delta E_{\,nl}^{\widetilde{\mathcal{O}}}(j)=-\frac{Z_{\text{\tiny{Pb}}}\,e^2\,b^2A_{j}}{2\tilde{\Lambda}_{\gamma}^{2j}\sin(j\pi)}&&\int d^{3}r~\lvert\Psi^{\mu^{-}}_{nl}(\vec{r})\rvert^{2}\int\frac{d^{3}\vec{q}}{(2\pi)^{3}}~e^{i\vec{q}\cdot \vec{r}}(\vec{q}^{\,2})^{-1}\tilde{f}(\vec{q})\int dx\,dy\,\delta(x+y-1)\,y^{1-j}\\ \nn
&&\frac{\left(M^{2}\right)^{j-1}}{(4\pi)^{2}}\left[\tilde{\Delta}(q^{2},\mu^2)~\frac{\log\tilde{\Delta}(q^{2},\mu^2)/M^{2}}{\left(1+\eta\right)}\right]
\eea
where $\Delta(q^{2},\mu^2)$ and $\tilde{\Delta}(q^{2},\mu^2)$ are as defined earlier in section II. 
\par
The energy shift in the $1S-2P$ transition in muonic lead may now be calculated  as 
\ba
 \Delta E^{\,\mathcal{O}}_{2p-1s}&=&\delta E^{\,\mathcal{O}}_{2p}-\delta E^{\,\mathcal{O}}_{1s}\\ \nn
 \Delta E^{\,\tilde{\mathcal{O}}}_{2p-1s}&=&\delta E^{\,\tilde{\mathcal{O}}}_{2p}-\delta E^{\,\tilde{\mathcal{O}}}_{1s}
\ea
\begin{figure}
\begin{center}
\includegraphics[width=7.5cm,angle=-90]{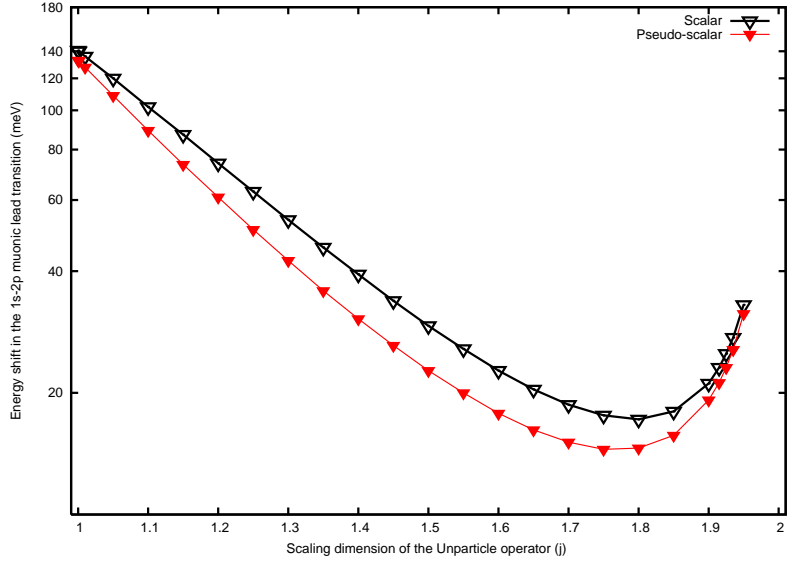}
\end{center}
\caption{Estimates on the magnitude of the energy shift $\lvert\delta E_{\,2p}^{\mathcal{U}}-\delta E_{\,1s}^{\mathcal{U}}\rvert$ due to possible scalar/pseudo-scalar Unparticle vacuum polarizations. We have taken $\mu\simeq\,0$, $b=c\sim\,\mathcal{O}(1)$ and $\Lambda_{\gamma},\, \tilde{\Lambda}_{\gamma}\sim\,246\,\text{GeV}$ as the reference value at the renormalization scale $M\simeq\Lambda_{\mathcal{U}}\simeq\,\Lambda_{\gamma},\, \tilde{\Lambda}_{\gamma}$. As we have mentioned previously the quoted values for the Uehling shifts should be interpreted as being accurate only up to $\mathcal{O}(1)$ factors to accommodate suitable ranges for the interaction scales and other approximations. It is clear from the plot that the Uehling shifts from Eqs.\,(\ref{seshift}) and (\ref{pseshift}) can under very general scenarios be in the ball-park of a few times 0.1 eV.}
\label{Eshift}
\end{figure}
using Eqs.\,(\ref{seshift}) and (\ref{pseshift}) with the nuclear profile of Eq.\,(\ref{nucp}). The results of the numerical computation, for $\mu\rightarrow0$, are shown in Fig.\,\ref{Eshift}. We observe from the plot that the energy shift due to pseudo-scalar Unparticle vacuum polarization is generally smaller in magnitude than the corresponding scalar case assuming other parameters are identical. This is probably an artifact of our approximation in Eq.\,(\ref{prescp}) as we commented previously. As one approaches $j\rightarrow\, 1^{+}$ though the energy shifts due to scalar and pseudo-scalar cases start to approach the same value of $\Delta E^{\,\mathcal{U}}_{2p-1s}\simeq\,0.14\,\text{eV}$ as expected. Note also that in the $j\rightarrow\, 1^{+}$ limit the ordinary scalar/pseudo-scalar case is recovered in a continuous manner.
 \par
Therefore, it may be claimed that for $\mu\rightarrow 0$ the typical shift in the $1S-2P$ muonic lead transition due to the Unparticle Uehling potential is in the $\mathcal{O}(0.1)$ eV range with reasonable assumptions about the parameters in the theory. This energy shift, if it exists, is similar in magnitude to QED corrections from the virtual Delbr\"uck effect (light-by-light scattering), to higher angular momentum transitions, in $\mu^{-}-\text{Pb}^{\,208}_{\,82}$ ~\cite{Borie:1982ax}. For example the virtual Delbr\"uck effect contributes
\ba
\Delta E^{\text{\tiny{Delb.}}}_{6h-5g}\simeq~0.4\,\text{eV}
\ea
in $\mu^{-}-\text{Pb}^{\,208}_{\,82}$. The Unparticle induced energy shift may also be compared to the vacuum polarization contribution in the hydrogenic Lamb shift which is in the $10^{-7}$ eV range (ordinary hydrogen) or eV range (muonic hydrogen). From (\ref{thexpdisc}) we see nevertheless that the discrepancy between measurement and theory of the $2p_{1/2}-1s_{1/2}$ and $2p_{3/2}-1s_{1/2}$ muonic lead transitions is in the range of many eV~\cite{{Bergem:1988zz},{Kessler:1975ju}}. This means that the estimated energy shift is about $10^2-10^3$ times smaller than the discrepancies quoted in (\ref{thexpdisc}). We will analyze a more realistic scenario with $\mu\neq 0$ later and will note then that the main features are not drastically modified from the simplistic $\mu\rightarrow 0$ case. 
\par
It is also seen from the plot that the energy-shift is \textit{very nearly linear}, with respect to the scaling dimension $j$, for values of $j$ close to $1$. In fact it may be verified analytically from Eqs.\,(\ref{eshiftscalar}), (\ref{eshiftpscalar}), (\ref{seshift}) and (\ref{pseshift}) that for $j=1+\eta$ with $\eta\ll\,1$ we may expand
\ba
\delta E_{\,nl}^{\Phi}-\delta E_{\,nl}^{\mathcal{U}}(j)\sim~C_{1}\,\eta+C_{2}\,\eta^{2}+C_{3}\,\eta^{3}+\hdots
\ea
where dimensionful factors have been omitted and $C_{n}$ are numerical coefficients. This implies that very close to $j=1$ we have to leading order 
\ba
\delta E_{\,nl}^{\phi}-\delta E_{\,nl}^{\mathcal{U}}(j)\propto~\eta
\ea 
as suggested by the plot and Eq.\,(\ref{prescp}). The above expression indicates that close to $j=1$ the additional energy shift in the $1S-2P$ transition compared to the ordinary scalar/pseudo-scalar case comes from the fractional scaling dimension part ($\eta$) of the Unparticle. This seems very satisfactory in the sense that any correction to the ordinary scalar/pseudo-scalar case is just proportional to the ``Unscalarness" of the Unparticle. This seems to be in the spirit of deconstruction~\cite{Stephanov:2007ry} and is similar to the observation in~\cite{Cacciapaglia:2007jq} 
\ba
\sigma_{\mathcal{U}}(j)~=~(2-j)~\sigma_{\phi}(j\rightarrow 1)
\ea
for the Unparticle production cross-section when the Unparticle sector is gauged under the SM.
\par
One also observes from Fig.\,\ref{Eshift} the onset of singular behavior as $j\rightarrow\,2$ as discussed in section I. As pointed out this is a pathology arising from the fact that close to $j=2$ the model becomes more and more UV sensitive. One may try to mitigate this singular behavior near $j=2$ by adding local contact terms~\cite{Grinstein:2008qk}.
\par
Concerning the sign of the Unparticle energy shifts it is inferred that there are no differences between the scalar and pseudo-scalar Unparticles. It found from our expressions in (\ref{seshift}) and (\ref{pseshift}) that the scalar as well as the pseudo-scalar Unparticle corrections to the $1S-2P$ transitions are \textit{positive} 
\bea
\Delta E^{\text{\,\tiny{ total}}}_{2p-1s}&\simeq&\Delta E^{\text{\tiny{ SM}}}_{2p-1s} +\lvert \Delta E^{\,\mathcal{O}}_{2p-1s}\rvert\\ \nn
\Delta E^{'\,\text{\tiny{ total}}}_{2p-1s}&\simeq&\Delta E^{\text{\tiny{ SM}}}_{2p-1s} +\lvert \Delta E^{\,\tilde{\mathcal{O}}}_{2p-1s}\rvert
\eea
\par
In the present case, for $\mu\rightarrow 0$, from Eqs.\,(\ref{lapavgSPS}) and (\ref{SPSenghier}) we would expect the level structure of the Uehling shifts to be ordered as
\ba
\Delta E^{\,\mathcal{U}}_{2s-1s}~&<&~\Delta E^{\,\mathcal{U}}_{2p-1s}\\
\Delta E^{\,\mathcal{U}}_{3s-1s}~&<&~\Delta E^{\,\mathcal{U}}_{3p-1s}~<~\Delta E^{\,\mathcal{U}}_{3d-1s}
\ea
These expectations from Eqs.\,(\ref{lapavgSPS}) and (\ref{SPSenghier}) are confirmed by numerical calculations. The variations between the $l=0,\,1,\,\text{and }2$ states are generally found to be of the order of a few $0.01\,\text{meV}$  for $n=3$ and of the order of a few meV for $n=2$.
\par
Let us now turn our attention to the case of \textit{broken scale invariance} where $\mu\neq0$. Even in the case of a fermiophobic Unparticle an effective interaction term with Higgs fields of the form
\ba
\Lambda^{2-j}_{h}~\mathcal{U}\,\mathcal{H}^{\,2}
\ea
is expected to be present. Here $\mathcal{H}$ stands for a generic Higgs field. As pointed out in the literature this interaction term is unique~\cite{{Fox:2007sy},{Strassler:2006im}} in the sense that it is a super-renormalizable term (since $1\leq j<2$ from Mack's unitarity). Even if one starts with a near zero coupling of this term, at a high energy, it will be generated through renormalization group flow and will break the conformal invariance of the Unparticle sector, at a scale $\mu$, when the Higgs develops a vacuum expectation value. Thus there is a very strong theoretical reason to expect scale invariance, in the fermiophobic Unparticle sector, to be broken at some scale after electroweak symmetry breaking. 
\par
Moreover, from the observational viewpoint, it is expected that there should be very stringent constraints on massless/light scalar degrees of freedom from cosmology and astrophysics. For a scalar/pseudo-scalar Unparticle coupling to two photons  with $\mu\rightarrow 0$ (i.e. perfect scale invariance) there are very strong bounds on the couplings from \textit{supernovae cooling} \cite{{Freitas:2007ip},{Das:2007nu}, {Davoudiasl:2007jr}} for example. Also, in the case of perfect scale invariance, there are very tight bounds on the couplings from \textit{big-bang nucleosynthesis} (BBN)~\cite{{Davoudiasl:2007jr}, {McDonald:2008uh}}. These constraints effectively render any collider or low-energy experiments ineffective in probing the Unparticle sector. 
\par
We would therefore like to estimate a lower bound on the scale invariance breaking effective mass ($\mu$), for the fermiophobic Unparticle, that would let us evade some of these constraints. So, assuming $\mu\neq 0$, let us first look at the typical Primakoff process with a fermiophobic Unparticle that could potentially contribute to supernovae (SN) cooling 
\ba
\gamma(k_{1})~+~\gamma(k_{2})\xrightarrow{\mu\neq0}~\mathcal{O}(p)
\ea
Here $\mathcal{O}$ is a fermiophobic scalar Unparticle and one of the photons is sometimes assumed to be off-shell. We are only interested in estimating a lower bound for the scale breaking parameter $\mu$ that would not violate supernovae constraints. Thus for simplicity we assume that the photons are transverse and ignore any plasmon effects (where the photon gets a longitudinal polarization in the plasma). Using
\ba
\vert\langle0\vert\mathcal{O}_{\mathcal{U}}(0)\vert P\rangle\vert^{2}\,\rho(P^{2})=~A_{j}\,\Theta(P^{0})\,\Theta(P^{2}-\mu^{2})\,\left(P^{2}-\mu^{2}\right)^{j-2}
\ea
and the Feynman rule from Eq.\,(\ref{slag}) we may compute the invariant amplitude to get
\ba
\lvert\mathcal{M}_{\mu\neq 0}({\gamma+\gamma\rightarrow\mathcal{O}})\rvert^{2}=\frac{c^{2}\,A_{j}}{4\,\Lambda_{\gamma}^{2j}}~\Theta(p^{0})~\Theta(s-\mu^{2})~s^{2}~(s-\mu^{2})^{j-2}
\ea
where $s=(k_{1}+k_{2})^{2}=p^{2}$ and $\Theta$ is the Heaviside function. From the above amplitude we may estimate the Unparticle Primakoff cross-section to be
\ba
\sigma_{\mu\neq0}({\gamma+\gamma\rightarrow\mathcal{O}})\simeq~\frac{c^{2}\,A_{j}}{8\,\Lambda_{\gamma}^{2j}}~s~(s-\mu^{2})^{j-2}
\ea
Using the above expression, the emissivity may be calculated from
\ba
\dot{\epsilon}_{\mu\neq 0}(\gamma+\gamma\rightarrow\mathcal{O})\simeq\frac{\langle n^{1}_{\gamma}\,n^{2}_{\gamma}~\sigma_{\mu\neq0}({\gamma+\gamma\rightarrow\mathcal{U}})~v_{\text{\tiny{rel.}}}\,E_{\text{\tiny{CM}}}\rangle}{\rho_{\text{\tiny{SN}}}}
\ea
and gives
\bea
\dot{\epsilon}_{\mu\neq 0}(\gamma+\gamma\rightarrow\mathcal{O})&\simeq&\frac{1}{\rho_{\text{\tiny{SN}}}}\int\frac{d^{3}\vec{k_{1}}}{(2\pi)^{3}}\,\frac{2}{e^{\frac{E_{1}}{k\,T_{\text{\tiny{SN}}}}}-1}\int\frac{d^{3}\vec{k_{2}}}{(2\pi)^{3}}\,\frac{2}{e^{\frac{E_{2}}{k\,T_{\text{\tiny{SN}}}}}-1}\\ \nn
&&~\frac{s(E_{1}+E_{2})}{2\,E_{1}E_{2}}~\frac{c^{2}\,A_{j}}{8\,\Lambda_{\gamma}^{2j}}~s~(s-\mu^{2})^{j-2}
\eea
where $\rho_{\text{\tiny{SN}}}$ is the average supernovae core density and $T_{\text{\tiny{SN}}}$ is the mean core temperature in the supernovae.  We define in the usual way the dimensionless variables
$x_{1}=E_{1}/kT_{\text{\tiny{SN}}}$ and $x_{2}=E_{2}/kT_{\text{\tiny{SN}}}$ to write the emissivity as
\be
\dot{\epsilon}_{\mu\neq 0}(\gamma+\gamma\rightarrow\mathcal{O})\simeq\frac{c^{2}\,A_{j}\,T^{2j+5}_{\text{\tiny{SN}}}}{16\pi^{4}\Lambda^{2j}_{\gamma}\rho_{\text{\tiny{SN}}}}~\int_{0}^{\infty}\,dx_{1}\int_{\frac{\mu^{2}}{4\,T^{\,2}_{\text{\tiny{SN}}}x_{1}}}^{\infty}\,dx_{2}~\frac{x_{1}\,x_{2}\,(x_{1}+x_{2})^{2j+1}}{(e^{x_{1}}-1)(e^{x_{2}}-1)}~\left(1-\frac{\mu^{2}}{T^{\,2}_{\text{\tiny{SN}}}(x_{1}+x_{2})^{2}}\right)^{j-2}
\label{ggusn}
\eq
In the limit of perfect scale invariance, $\mu\rightarrow 0$, we recover from the above the expressions in \cite{{Freitas:2007ip},{Das:2007nu}}. Eq.\,(\ref{ggusn}) may be compared with the fermiophilic case of a \textit{vector} Unparticle ($\mathcal{U}_{V}$), with $\mu\neq 0$, coupling to fermions. In this case a process such as $\nu\nu\rightarrow\,\mathcal{U}_{V}$ in the supernovae core can lead to cooling and the emissivity is given by~\cite{Barger:2008jt}
\ba
\dot{\epsilon}_{\mu\neq 0}(\nu+\nu\rightarrow\mathcal{U}_{V})&\simeq&\frac{g_{\nu}^{2}\,A_{j}\,T^{2j+3}_{\text{\tiny{SN}}}}{16\pi^{4}\Lambda^{2j-2}_{\psi}\rho_{\text{\tiny{SN}}}}~\int_{0}^{\infty}\,dx_{1}\int_{\frac{\mu_{V}^{2}}{4\,T^{\,2}_{\text{\tiny{SN}}}x_{1}}}^{\infty}\,dx_{2}~\frac{(4\,x_{1}\,x_{2})^{j}\,(x_{1}+x_{2})}{(e^{x_{1}}+1)(e^{x_{2}}+1)}~\left(1-\frac{\mu_{V}^{2}}{4\,T^{\,2}_{\text{\tiny{SN}}}x_{1}\,x_{2}}\right)^{j-2}
\ea
Note that the power law dependence of $T_{\text{\tiny{SN}}}$ on the scaling dimension $j$ is very different, between the fermiophobic and fermiophilic cases, even if one assumes that the other parameters $\mu\simeq\mu_{V}$, $\Lambda_{\gamma}\simeq\Lambda_{\psi}$ and $\mathcal{O}(1)$ coefficients $c\simeq g_{\nu}$. But we shall see below that the bound on the scale breaking parameter comes out to be of the same order of magnitude in both the cases due to the dominance of the Boltzmann suppression factor.
\par
To be consistent with supernovae models and constraints from SN1987A we require~\cite{Raffelt:1990yz}
\ba
\dot{\epsilon}_{\text{\tiny{SN}}}\lesssim~10^{15}~\text{J/Kg. s}
\ea
Eq.\,(\ref{ggusn}) may be integrated and expressed in a compact form in the limit of $\mu\gg T_{\text{\tiny{SN}}}$ with the Boltzmann suppression term factored out. An explicit numerical computation with the standard values
\ba
T_{\text{\tiny{SN}}}~\simeq~30\,\text{MeV}~~~;~~~~~~\rho_{\text{\tiny{SN}}}~\simeq~10^{18}~\text{Kg}/\text{m}^{3}
\ea
 and the above emissivity criterion gives 
\be
\mu~\gtrsim~1.25~\text{GeV}
\label{mufermp}
\eq 
This lower bound for the effective mass, of a  fermiophobic scalar/pseudo-scalar Unparticle coupling only to photons, is found to be very close to the result in the fermiophilic case $\nu+\nu\rightarrow\mathcal{U}_{V}$ where it was found that~\cite{Barger:2008jt}
\ba
\mu_{\,V}~\gtrsim~1~\text{GeV}
\ea
\par
Based on the bound in Eq.\,(\ref{mufermp}) we may take a minimal value for the effective Unparticle mass to be $\mu\simeq\,2\,\text{GeV}$. This choice already leads to an emissivity well below the allowed limit. A crude upper bound on the scale breaking parameter is given by the requirement that $\mu\ll\Lambda_{\mathcal{U}}$. Since we have adopted the prejudice that the Unparticle scale is probably close to the electroweak-symmetry breaking scale this implies that in our model we are assuming $\mu\ll\,\mathcal{O}(1)\,\text{TeV}$. 
\par
Let us now turn our attention to some of the constraints from cosmology. For the fermiophobic Unparticle sector not to interfere with BBN we must require that the Unparticles stay decoupled during that epoch~\cite{Davoudiasl:2007jr}. In the case of a fermiophobic Unparticle with $\mu\rightarrow 0$ the condition $\Gamma_{\text{\tiny{SM}}\rightarrow\mathcal{O}}<H$ (during the radiation dominated era), by simple dimensional analysis, becomes
\ba
\Gamma_{\gamma\gamma\mathcal{O}}\sim~\frac{c^{2}}{\Lambda^{2j}_{\gamma}}~T^{\,2j+1}~<~\left(\frac{T^{\,2}}{10^{18}\,\text{GeV}}\right)
\ea
This implies that in the range $j\in[1,2)$ we are considering the rate $\Gamma_{\gamma\gamma\mathcal{O}}$ \textit{red-shifts faster} than the Hubble parameter ($H$). It is required that the fermiophobic Unparticles decouple before BBN and not get reheated during SM phase transitions to satisfy $\rho_{\mathcal{U}}\ll\,\rho_{\text{\tiny{SM}}}$. This may be achieved by requiring that decoupling happen before the QCD phase transition at an energy $T\gtrsim\,1\,\text{GeV}$~\cite{Davoudiasl:2007jr}. Note that a fermiophobic Unparticle sector, coupling to photons, does not generally re-couple after BBN with SM fields because $\Gamma_{\gamma\gamma\mathcal{O}}$ red-shifts faster than the Hubble parameter. This may again be contrasted with the fermiophilic case of a vector Unparticle coupling to fermions through the effective operator
\ba
\frac{g_{f}}{\Lambda^{j-1}_{\psi}}~\bar{\psi}\,\gamma_{\mu}\,\psi\,\mathcal{O}^{\mu}_{V}
\ea
 where re-coupling is possible when $1\leq\,j\,\leq\,3/2$~\cite{Davoudiasl:2007jr}. 
 \par
 When scale invariance is broken ($\mu\neq0$), since $\Gamma_{\mathcal{O}}\sim\,n_{\text{\tiny{EQ}}}\,\langle\sigma\lvert v\rvert\rangle$, the relevant processes are Boltzmann suppressed by factors of $e^{-\mu/T}$ when $\mu>\,T$. Thus the BBN constraints can be evaded as long as $\mu$ is above $\Lambda_{\text{\tiny{QCD}}}$~\cite{Barger:2008jt}. We must point out that we have tacitly assumed that the fermiophobic Unparticles can decay into SM fields~\cite{Rajaraman:2008bc} when $\mu$ is sufficiently non-zero and are therefore not stable. There has been some discussion in the literature on this particular issue~\cite{{Stephanov:2007ry}, {McDonald:2008uh}, {Rajaraman:2008bc}}. All the arguments above suggest that while calculating the Uehling shifts with $\mu\neq0$ we must also consider the possibility that $\mu$ is higher than the minimal value of $2\,\text{GeV}$. Thus without loss of generality we will consider a range
\ba
2~\text{GeV}~\lesssim~\mu~\ll~M_{Z}\simeq\,91\,\text{GeV}
\ea
for both the scalar and pseudo-scalar Unparticles such that there is still a substantial conformal window $(\mu,\,\Lambda_{\mathcal{U}}]$. The above choice would also ensure that any modification to the gauge kinetic term ($\Delta\alpha^{-1}$) near the scale $\mu$ is within experimental limits~\cite{Bander:2007nd} and that the effects due to SM Higgs-Unparticle mixing, if present, are suppressed~\cite{Barger:2008jt}. We will demonstrate in a short while that the actual energy shifts are relatively insensitive to small shifts in the $\mu$ parameter.
\par
The pseudo-scalar Unparticle potential energies for two choices of the scale breaking parameter $\mu$ are shown in Fig.\,\ref{mupotPS}. This figure is to be compared with that in the case of perfect scale invariance illustrated in Fig.\,\ref{potPS}. It is noted that the potential in the $\mu\neq 0$ case has been suppressed by $\mathcal{O}(1)$ factors compared to the $\mu\rightarrow 0$ case. The general features of the Unparticle Uehling potential nevertheless remain unchanged between the $\mu\rightarrow 0$ and $\mu\neq 0$ scenarios. 
\begin{figure}
\includegraphics[width=8.75cm,angle=0]{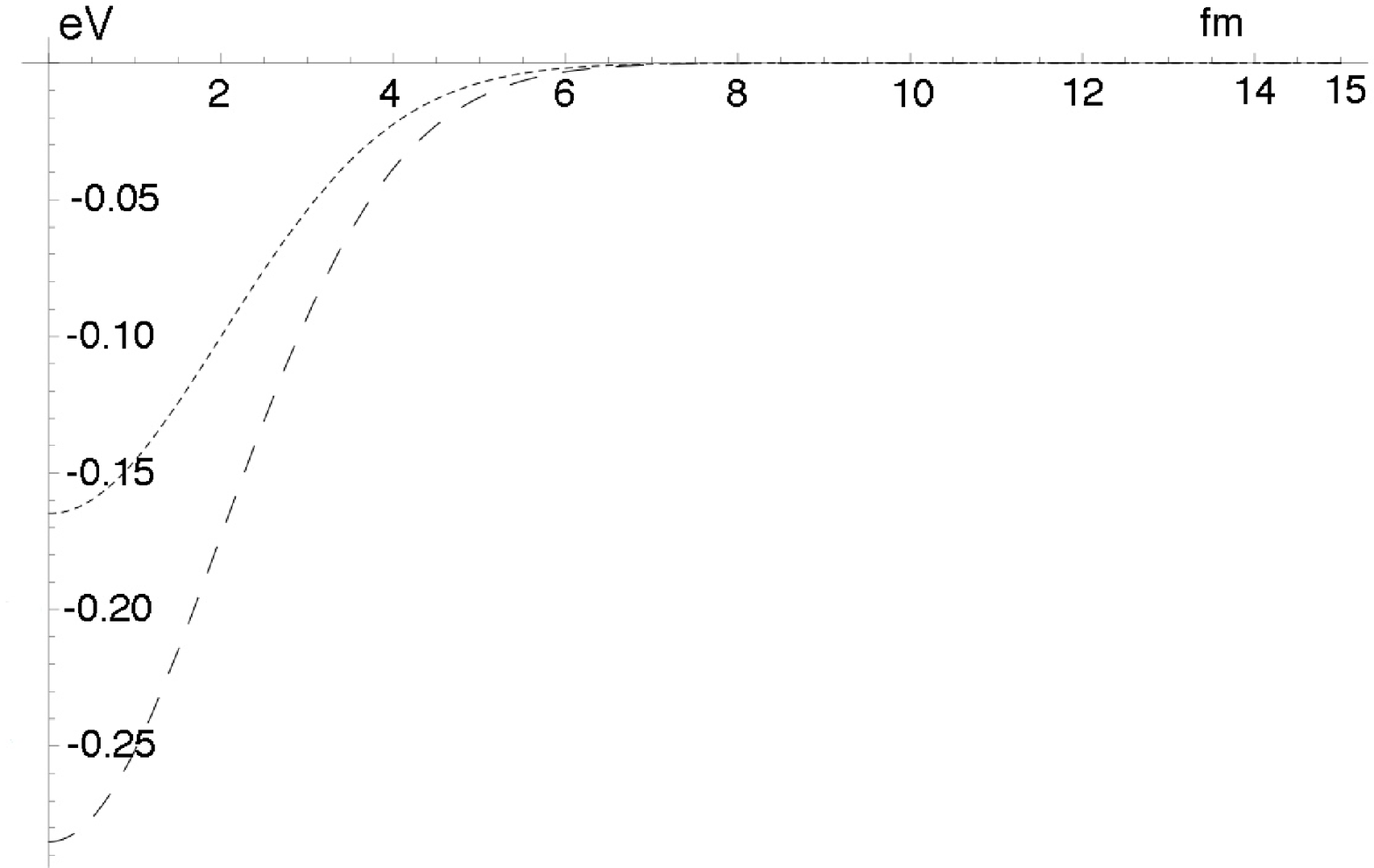}
\hfill
\includegraphics[width=8.75cm,angle=0]{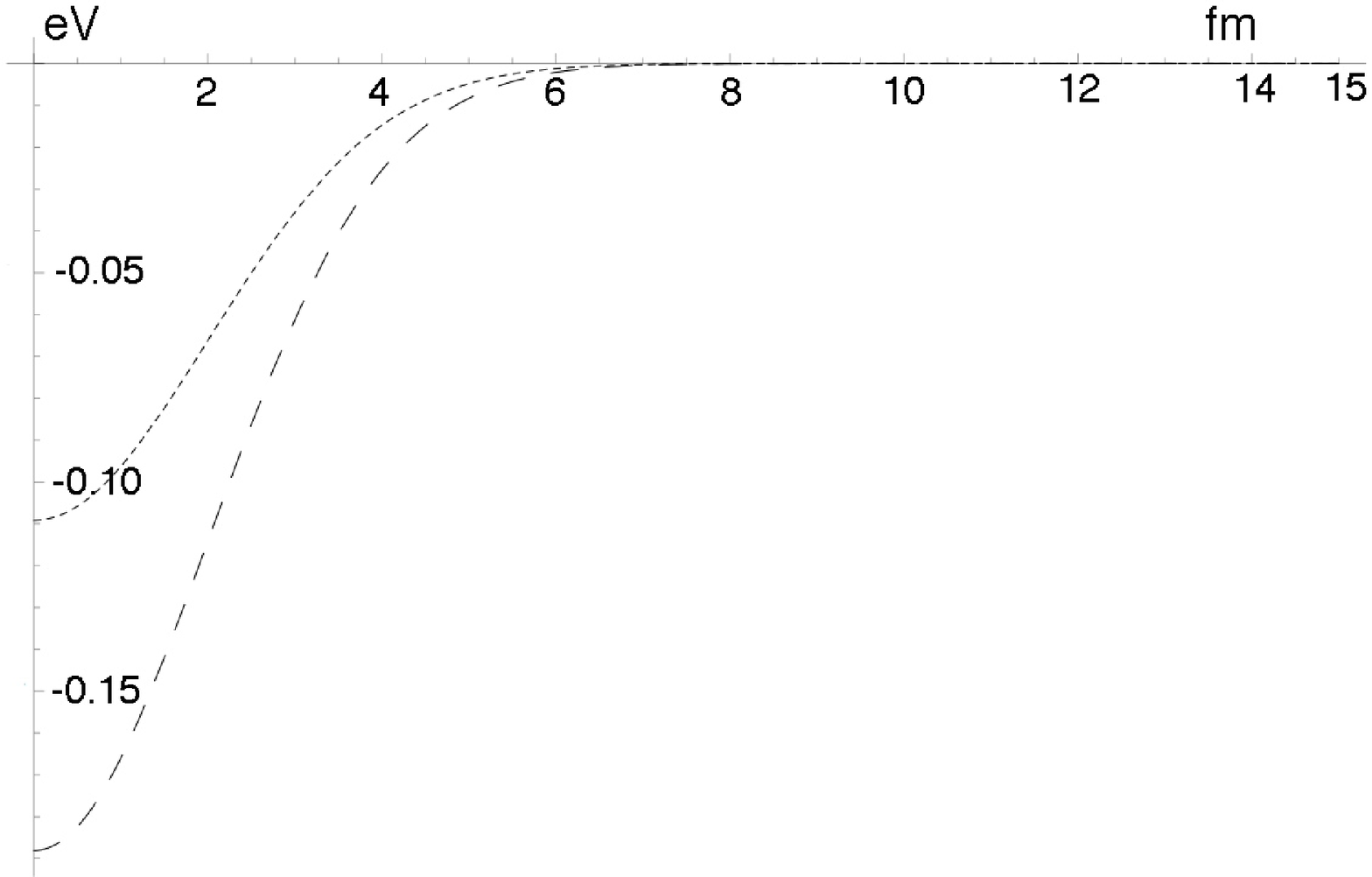}
\caption{The Unparticle Uehling potential energies for the pseudo-scalar case with the same parameters as  Fig.\,\ref{potPS} except assuming $\mu$ is $2\,\,\text{GeV}$ (left panel) and $10\,\,\text{GeV}$ (right panel). The $j=1.01$ (dashed line) and $1.15$ (small-dashed line) potential energies are again illustrated as functions of $r$. Note that the magnitude of the potential at any specific value of $r$ has been reduced in both cases compared to the case of perfect scale invariance $\mu\rightarrow 0$.}
\label{mupotPS}
\end{figure}
\par
Let us now calculate the energy shifts in the muonic lead transitions when $\mu\neq 0$. Consider the pseudo-scalar Unparticle with scale invariance broken. We may re-write Eq.\,(\ref{pseshift}) again by defining the dimensionless variable $z=\lvert\vec{q}\rvert\,r$. The expression becomes after simplification
\bea
\label{epsshiftmu}
\delta E_{\,nl}^{\widetilde{\mathcal{O}}}(j)&=&-\frac{Z_{\text{\tiny{Pb}}}\, e^2\, b^2\,A_{j}\,\mu^{2}}{64\pi^{4}\tilde{\Lambda}_{\gamma}^{2j}\sin(j\pi)}\int d^{3}r~\lvert\Psi^{\mu^{-}}_{nl}(\vec{r})\rvert^{2}\int_{0}^{1} dx\,\left(1-x\right)^{2-j} \frac{(M^{2})^{j-1}}{j\,r}\\ \nn
&&\int_{0}^{\infty}dz~\frac{\sin z}{z}~e^{-\zeta z^{2}/(2 r^{2})}\left[\left(1+\frac{x\,z^2}{\mu^2\,r^2}\right)\log\bigg\{1+\frac{x\,z^2}{\mu^2\,r^2}\bigg\}+\frac{x\,z^2}{\mu^2\,r^2}\log\bigg\{\frac{(1-x)\mu^2}{M^2}\bigg\}\right]
\eea
where we have made an over-subtraction at $q=0$ to get a consistent $q\rightarrow 0$ limit. From the above expression we expect that for any fixed value of the radial coordinate ($r$) the dominant contribution to the $z$-integral should come from the region of integration with
\be
z^{2}~\lesssim~\frac{2\,r^{2}}{\zeta}
\label{momentumcut}
\eq
or its vicinity. For the muonic lead system we are primarily interested in a suitable value of the nuclear charge density parameter $\zeta$, in Eq.\,(\ref{nucp}), was found to be $\zeta\simeq\,4\,\text{fm}^{2}$. We found that a minimal value of $\mu$ satisfying astrophysical constraints is  $\mu\simeq\,2\,\text{GeV}\equiv\,10.14\,\text{fm}^{-1}$. From these observations we note therefore that for a fixed value of $r$
\ba
x\,z^{2}~\lesssim~~\frac{2\,r^{2}}{\zeta}~\ll~\mu^{2}\,r^{2}
\ea
This means that in the relevant region of the parameter space we may expand the logarithm in Eq.\,(\ref{epsshiftmu}) as
\ba
\log\bigg\{1+\frac{x\,z^2}{\mu^2\,r^2}\bigg\}\simeq~\frac{x\,z^2}{\mu^2\,r^2}-~\frac{1}{2}\,\left(\frac{x\,z^2}{\mu^2\,r^2}\right)^{2}+~\ldots
\ea
to get an integral that is separable in $x$ and $z$ variables. Note that this approximation becomes more and more accurate as we raise the value of $\mu$. The variable separable integral may now be evaluated analytically or numerically to calculate the expression in Eq.\,(\ref{epsshiftmu}). The calculated energy shifts to the $1S-2P$ transition for various values of the scaling dimension $j$ and $\mu$ are shown in the left panel of Fig.\,\ref{Eshift_mu}. It is noted that compared to the $\mu\rightarrow 0$ case the energy shifts are smaller, assuming all other parameters remain the same, by $\mathcal{O}(1)$ factors. Thus we find that incorporating broken scale invariance with a non-zero value of $\mu$ does not seem to alter the energy shifts drastically from their $\mu\rightarrow 0$ values and the changes are only by whole number factors. Once again, maintaining continuity, the ordinary pseudo-scalar case is recovered as $j$ approaches $1$ due to Eq.\,(\ref{prescp}). If $\mu$ is very large and the corresponding energy shifts very small then the finite contributions from the higher order counter terms in the effective Lagrangian may become important and the numerical approximation we adopt, of keeping only the lowest order terms from (\ref{slag}) and (\ref{pslag}), may break down.  
\begin{figure}
\includegraphics[width=6.2cm,angle=-90]{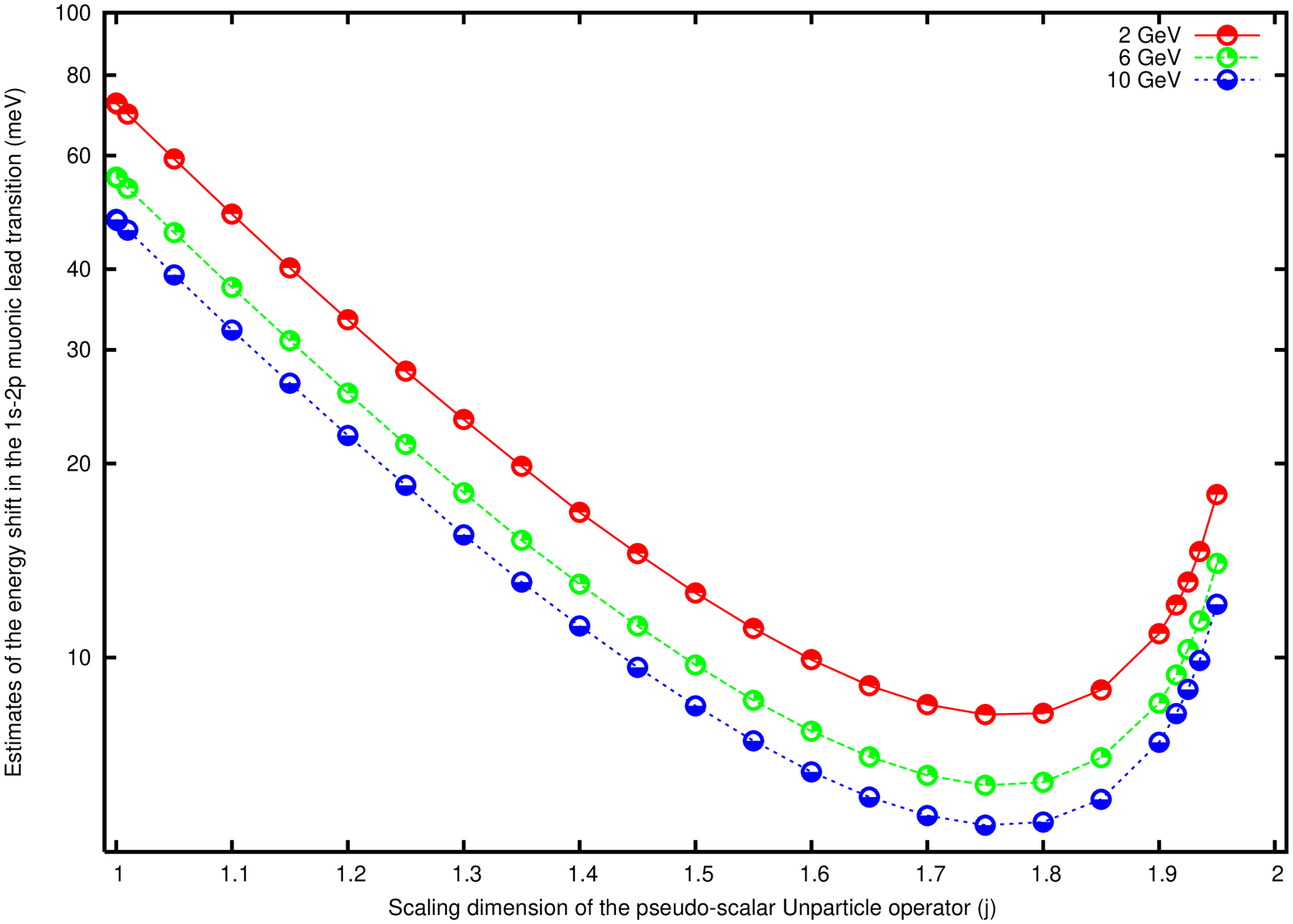}
\hfill
\includegraphics[width=6.2cm,angle=-90]{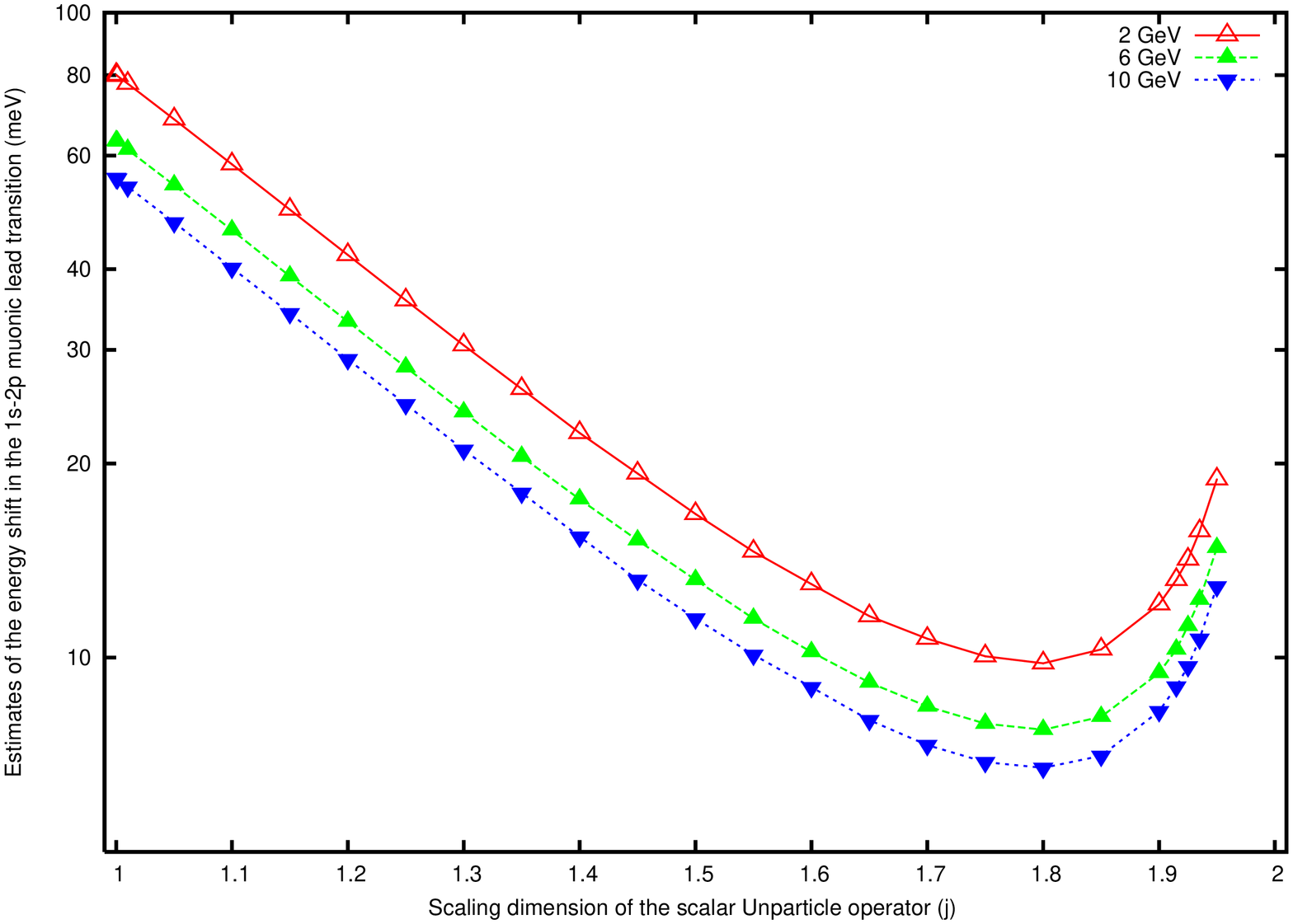}
\caption{Magnitude estimates of the energy shift in the $1S-2P$ muonic lead transition due to pseudo-scalar (left panel) and scalar (right panel) Unparticle vacuum polarizations. The plots are for $\mu\simeq\,2\,\text{GeV},~ 6\,\text{GeV}$ and $10\,\text{GeV} $. With these choices of $\mu$ there is still a substantial conformal window between $\mu$ and $\Lambda_{\mathcal{U}}$. The other parameters are taken to be the same as in Fig.\,\ref{Eshift}. Note that the energy shift is lower in both cases compared to the $\mu\rightarrow 0$ case, but not drastically. In fact it is observed that the energy shifts are relatively substantial even for $\mu\gg m_{\mu^{-}}\simeq 0.1\,\text{GeV}$ compared to $\mu\rightarrow 0$. We will explore the reasons for this insensitivity to $\mu$ shortly. In general it is observed that as we increase the effective mass the energy shift decreases. As before the energy shifts should be interpreted as accurate only up to undetermined $\mathcal{O}(1)$ factors.}
\label{Eshift_mu}
\end{figure}
\par
In the scalar Unparticle case the expression for the energy shift becomes 
\bea
\label{esshiftmu}
\delta E_{\,nl}^{\mathcal{O}}(j)&=&\frac{Z_{\text{\tiny{Pb}}}\, e^2\, c^2\,A_{j}\mu^{2}}{64\pi^{4}\Lambda_{\gamma}^{2j}\sin(j\pi)}\int d^{3}r~\lvert\Psi^{\mu^{-}}_{nl}(\vec{r})\rvert^{2}\int_{0}^{1} dx\,\left(1-x\right)^{2-j} \frac{(M^{2})^{j-1}}{j\,r}\int_{0}^{\infty}dz~\frac{\sin z}{z}~e^{-\zeta z^{2}/(2 r^{2})}\\ \nn
&& \left[\left(1+\,\left((1+j)x-j\right)\,\frac{z^2}{\mu^2r^2}\right)\log \bigg\{1+\frac{x z^2}{\mu^2 r^2}\bigg\}+\,\left((1+j)x-j\right)\frac{z^2}{\mu^2 r^2}\log\bigg\{\frac{(1-x)\mu^2}{M^2}\bigg\}\right]
\eea
after an over-subtraction at $q=0$. In the relevant region where $x\,z^{2}\lesssim 2\,r^{2}/\zeta\ll\mu^{2}\,r^{2}$ we may again simplify the integral and compute it to obtain the energy shifts. The calculated values are shown in Fig.\,\ref{Eshift_mu}, right panel, for various values of the $\mu$ parameter. The limit of $j\rightarrow1^{+}$, as before, corresponds to the ordinary scalar case. Again, the correction from the scalar Unparticle vacuum polarization is found to be larger in magnitude than the corresponding pseudo-scalar Unparticle case. But this may as before be an artifact of our approximation in Eq.\,(\ref{prescp}). The corrections are both \textit{positive} as in the $\mu\rightarrow 0$ case. 
\par
Once again we observe that the variation of the scalar Unparticle Uehling shift with $\mu$ is only by $\mathcal{O}(1)$ factors. This implies that the energy shift is relatively insensitive to changes of the scale breaking parameter, across a wide range, in both the scalar and pseudo-scalar cases. More specifically the Uehling shifts are relatively unchanged even for $\mu\gg m_{\mu^{-}}\simeq0.1\,\text{GeV}$. In this sense the Unparticle oblique correction is sensitive to a wide range of $\mu$ values within our approximations. Let us try to understand this a little better.
\par
Let us look at the expression
\ba
\int_{0}^{\infty}dz~\frac{\sin z}{z}~e^{-\zeta z^{2}/(2 r^{2})}\left[\frac{\mu^{2}}{r}\log\bigg\{1+\frac{xz^{2}}{\mu^{2}r^{2}}\bigg\}+\frac{xz^{2}}{r^{3}}\log\bigg\{(1-x)\left(\frac{\mu^{2}}{M^2}+\frac{xz^{2}}{M^{2}r^{2}}\right)\bigg\}\right]
\ea
from Eq.\,(\ref{epsshiftmu}) for the pseudo-scalar Unparticle that we have re-written to include all the dependences on $\mu$. The first term in the above expression tends to zero in the $\mu\rightarrow 0$ limit and is absent in the case of perfect scale invariance. Even in the $\mu\neq0$ limit we note that since $x\,z^{2}\lesssim 2\,r^{2}/\zeta\ll\mu^{2}\,r^{2}$ in the relevant region the first term is relatively suppressed. Thus the correction from the first term when $\mu\neq0$ is small, even though the corresponding term in the case of $\mu\rightarrow 0$ is completely absent. Now let us look at the second term in the above expression, specifically the $\mu^2/M^2$ factor inside the logarithm. Again that factor inside the logarithm is obviously absent when $\mu\rightarrow 0$. Since we have adopted the \textit{ansatz} that the Unparticle scale is in the vicinity of the electroweak scale we haveup to $\mathcal{O}(1)$ factors $M\simeq\,\Lambda_{\mathcal{U}}\sim\,v$. Thus at the level of our approximation for typical values of allowed $\mu$ we have $\mu^2/M^2\ll\,1$. This again implies that the correction in the $\mu\neq0$ case, with respect to the $\mu\rightarrow 0$ case, from the second term is not very drastic. 
\par
For the scalar Unparticle case the arguments proceed exactly as above for the relevant expression
\ba
\int_{0}^{\infty}dz~\frac{\sin z}{z}~e^{-\zeta z^{2}/(2 r^{2})}\left[\frac{\mu^{2}}{r}\log\bigg\{1+\frac{xz^{2}}{\mu^{2}r^{2}}\bigg\}+\frac{\left((1+j)x-j\right)z^{2}}{r^{3}}\log\bigg\{(1-x)\left(\frac{\mu^{2}}{M^2}+\frac{xz^{2}}{M^{2}r^{2}}\right)\bigg\}\right]
\ea
and once again we conclude that the corrections to the $\mu\rightarrow 0$ case from the additional factors are not very large. Thus the relative stability of the Uehling energy shifts to variations in the scale breaking parameter $\mu$ may be traced to the momentum cut-off imposed by $e^{-\zeta z^{2}/(2 r^{2})}$ leading to the observation in Eq.\,(\ref{momentumcut}) and the fact that $\mu\lesssim M\simeq\,\Lambda_{\mathcal{U}}\sim\,v$ for typical values.
\par
The typical values for the Uehling energy shift in Fig.\,\ref{Eshift_mu} are in the range $\mathcal{O}(0.1)\,\text{eV}-\mathcal{O}(0.01)\,\text{eV}$. The $\mathcal{O}(0.1)\,\text{eV}$ as we commented previously is comparable to the contribution from light-by-light scattering in QED. One may get a feel for the lower value $\mathcal{O}(0.01)\,\text{eV}$ in the range by noting that it is of the same order of magnitude as the corrections from the QED fourth-order Lamb shift, to higher angular momentum states in $\mu^{-}-\text{Pb}^{\,208}_{\,82}$~\cite{Borie:1982ax}. For higher angular momentum states the muon's anomalous magnetic moment induces an additional  spin-orbit interaction in the muonic atom and this contributes to an energy shift. For instance in the $\mu^{-}-\text{Pb}^{\,208}_{\,82}$ $5g-4f$ transition the leading contribution to the fourth-order Lamb shift at order $\alpha^{2}\,(Z\alpha)$ was estimated to be~\cite{Borie:1982ax}
\ba
\Delta E^{\text{\tiny{\,4-LS}}}_{5g-4f}\simeq~0.025\,\text{eV}
\ea
The values of the Uehling shifts in Fig.\,\ref{Eshift_mu} may again be compared to the uncertainties in the precision measurements of $2p_{1/2}-1s_{1/2}$ and $2p_{3/2}-1s_{1/2}$ transitions in Table IV and the discrepancy between theory and measurement for these transitions in (\ref{thexpdisc}). It is noted that for typical values of the model parameters ($j$, $\Lambda_{\gamma}$, $c$ and $\mu$) the Uehling shift is again about a factor of $10^3-10^4$ below the values in (\ref{thexpdisc}). 
\par
We may also calculate the Uehling shifts for the low-lying $l=1$ and $l=2$ states with respect to $1S$. It is observed that for $\mu\neq0$ the Unparticle scalar and pseudo-scalar corrections again follow a hierarchy
\ba
\Delta E^{\,\mathcal{U}}_{2s-1s}~&<&~\Delta E^{\,\mathcal{U}}_{2p-1s}\\
\Delta E^{\,\mathcal{U}}_{3s-1s}~&<&~\Delta E^{\,\mathcal{U}}_{3p-1s}~<~\Delta E^{\,\mathcal{U}}_{3d-1s}
\ea 
consistent with our expectations in Eq.\,(\ref{SPSenghier}), for the case of perfect scale invariance, and also with our computations for the $\mu\rightarrow 0$ case before. We conclude that the choice of $\mu\neq0$ does not change the level structure of the Uehling shifts. Moreover the variations are generally of the same order of magnitude as in the $\mu\rightarrow 0$ case. For example, assuming $\mu\simeq\,2\,\text{GeV}$ and $\Lambda_{\mathcal{U}}\sim\,v$, the variations between the $l=0\,\text{and}\,1$ states for $n=2$ are again of the order of a few meV and for $n=3$ of the order of a few $0.01\,\text{meV}$. 
\par
 A speculation is that one may do precision spectroscopy of the low-lying atomic states in parallel with proposed efforts to observe \textit{coherent muon-electron conversion} in muonic atoms (see \cite{Czarnecki:1998iz} and references therein). The relevant process in coherent muon-electron conversion is
\ba
\mu^{-}~+~N~\rightarrow~e^{-}~+~N
\ea
where $N$ is a nucleon. It is believed that probes of resonant muon-electron conversion near a nucleus may be able to achieve a higher sensitivity to lepton-flavor-violation (LFV) compared to direct conversions~\cite{Ankenbrandt:2006zu}
\ba
\mu^{-}\rightarrow\,e^{-}+\gamma
\ea
\par
This opens the possibility that one may also perform precision spectroscopy on low-lying muonic atom states in these forthcoming experiments and reduce some of the discrepancies in Eq.\,(\ref{thexpdisc}). But reducing the discrepancy to a level of $\mathcal{O}(0.01)\,\text{eV}$ looks very improbable to us. 
\par
Since muon-conversion is a coherent process one might expect that the probability of muon-conversion in an atom $X^{A}_{Z}$ would go like $\sim\,Z^2$ (or $\sim\,A^2$). So when normalized to the muon-capture cross section we have heuristically
\ba
R_{\mu e}=~\frac{\Gamma\left[\mu^{-}+(A,Z)\rightarrow e^{-}+(A,Z)\right]}{\Gamma\left[\mu^{-}+(A,Z)\rightarrow \nu_{\mu}+(A,Z-1)\right]}\sim~\frac{Z^2}{Z}\approx~Z
\ea
while for the Uehling shift we are interested in, as we noted previously, the dependence on the atomic number $Z$ goes like
\ba
\frac{\delta E^{\text{\tiny{VP}}}_{nl}}{E_{\text{\tiny{Bound}}}}\sim~Z^{2}
\ea	
\par
It is not clear that the muonic lead system, with $Z=82$, that we are considering is suitable for the LFV measurements since it has been shown through detailed calculations~\cite{Kitano:2002mt} that for coherent muon-electron conversion the most ideal range is $Z\in[30,\,60]$. Thus it would be interesting, as we have previously mentioned, to explore the possibility of measuring the scalar/pseudo-scalar Uehling shifts in other atomic systems, with an intermediate $Z$ value, where the energy shifts may still be substantial while the system is also of interest to coherent muon-electron conversion experiments. It would also be interesting to estimate the Unparticle vacuum polarization effects in muonic atoms where high-precision LASER spectroscopy might be possible~\cite{Kawall:1997ei}. We hope that this work will be a modest pointer in this direction.
\par
It is of course an open possibility that if for some reason the interaction energy scales $(\Lambda_{\gamma},\,\tilde{\Lambda}_{\gamma})$ are too large ($\gg \mathcal{O}(1)\,\text{TeV}$) or the coefficients $(b,\,c)$ in Eqs.\,(\ref{slag}) and (\ref{pslag}) are very small the energy shifts will be even more suppressed and the Unparticle Uehling shift, even if it exists, will be nearly impossible to detect. Even under optimistic assumptions about the model parameters it is possible that the theoretical difficulties in calculating the required higher-order QED/nuclear effects in muonic atoms may be insurmountable or that obtaining precision spectra of the low-lying states is very difficult. In such a scenario the only hope would be to look for fermiophobic Unparticles in very high energy colliders or other systems that are more sensitive to fermiophobic Unparticles.  
\end{section}


\begin{section}{Summary}
In this work we tried to study some of the probable effects of a fermiophobic scalar/pseudo-scalar sector on bound state energy levels, specifically low-lying muonic atom levels, as a consequence of oblique corrections to the photon propagator. 
\par
Considering the scalar and pseudo-scalar fields to be Unparticle operators, without loss of generality, we examined the functional forms of the vacuum polarization functions and the induced Uehling potentials. Some interesting theoretical observations were made on the singular nature of the Unparticle induced Uehling potential and the behavior of the energy shifts in the limit of the scaling dimension approaching unity.
\par
It was estimated that for an Unparticle scale near the scale of electroweak symmetry breaking, in the low TeV range, the energy shifts in the low-lying muonic lead transitions could typically be of the order of a few 0.1 eV to a few 0.01 eV for some natural values of the model parameters. It was also pointed out that these magnitudes are comparable to bound state QED corrections, to the higher orbital angular momentum transitions in muonic lead, from the virtual Delbr\"uck effect (light-by-light scattering) and the fourth order Lamb-shift (at order $\alpha^{2}(Z\alpha)$) respectively. These conclusions are relatively unchanged even when one incorporates a breaking of the scale invariance by introducing an effective Unparticle mass $\mu$. 
\par
But the current discrepancy between muonic-lead spectroscopy and theory, especially nuclear theory, makes an interpretation of the Unparticle Uehling shift, if it really exists, extremely challenging. A conservative estimate is that such an interpretation would require an improvement in the discrepancy between theory and experiment, from about 20 years back, by a factor of  $1000-10000$. The recent, partial resolution of the long standing discrepancy in the $\Delta 2p$ and $\Delta 3 p$ NP calculations with results from muonic lead spectroscopy~\cite{{Bergem:1988zz},{Kessler:1975ju}} by A. Haga and co-workers~\cite{Haga:2007mx} is a promising step in this direction.
\par
We also mentioned that in cases where the UV sector has a very large fermion \textit{multiplicity} the above contribution may be greatly enhanced, but appealing to arguments of \textit{naturalness} we do not think this to be very plausible. But many interesting models being considered today (for example string-inspired QCD-like models~\cite{Sakai:2004cn}) allow for the possibility of a large fermionic sector and this perhaps beseeches us not to discard the possibility of fermiophobic Unparticle oblique corrections prematurely. 
\par
The other interesting direction is to consider intermediate-$Z$ muonic atoms where the nuclear uncertainties may be much better controlled while at the same time have sufficient fermiophobic Unparticle contributions to muonic-atom transitions, by virtue of the $\delta E^{\text{\tiny{VP}}_{nl}}/E_{\text{\tiny{Bound}}}\sim~Z^{2}$ enhancement. This is left for future work.
\par
In the context of the present study, of a possible fermiophobic Unparticle scalar/pseudo-scalar sector, we also briefly considered constraints from astrophysics and cosmology and put bounds on the fermiophobic Unparticle effective masses. Finally we speculated on improving muonic lead spectroscopy and theory in the context of forthcoming experiments that will study coherent muon-electron conversion. 
\end{section}


\begin{acknowledgments}
 I would like to thank Jonathan L. Rosner for many useful suggestions during the present study and a careful reading of the manuscript. I also thank D. Erkal, S. Farkas, D. McKeen, J. Galloway, D. Krohn and P. Draper for discussions. The author also acknowledges interesting comments from the referee. This work was supported in part by the United States Department of Energy under Grant No. DE-FG02-90ER40560.
\end{acknowledgments}


\end{document}